\documentclass[iop, revtex4]{emulateapj}
\usepackage{natbib,graphicx,amsmath,amsthm,ulem, subfigure}

\shorttitle{TRENDS IV: The Occurrence Rate of Giant Planets around M-Dwarfs}
\shortauthors{Montet et al.}

\begin{document}

\title{The TRENDS High-Contrast Imaging Survey. IV. \\ The Occurrence Rate of Giant Planets around M-Dwarfs}
\author{Benjamin T. Montet\altaffilmark{1}, Justin R. Crepp\altaffilmark{2}, John Asher Johnson\altaffilmark{3}, Andrew W. Howard\altaffilmark{4}, Geoffrey W. Marcy\altaffilmark{5}}
\email{btm@astro.caltech.edu}
\altaffiltext{1}{Cahill Center for Astronomy and Astrophysics, California Institute of Technology, 1200 E. California Blvd., MC 249-17, Pasadena, CA 91125, USA}
\altaffiltext{2}{Department of Physics, University of Notre Dame, 225 Nieuwland Science Hall, Notre Dame, IN 46656, USA}
\altaffiltext{3}{Harvard-Smithsonian Center for Astrophysics, 60 Garden Street, Cambridge, MA 02138, USA}
\altaffiltext{4}{Institute for Astronomy, University of Hawaii, 2680 Woodlawn Drive, Honolulu, HI 96822, USA}
\altaffiltext{5}{B-20 Hearst Field Annex, Astronomy Department, University of California, Berkeley, Berkeley, CA 94720, USA}

\begin{abstract}

Doppler-based planet surveys have discovered numerous giant planets but are incomplete beyond several AU. At larger star-planet separations direct planet detection through high-contrast imaging has proven successful, but this technique is sensitive only to young planets and characterization relies upon theoretical evolution models. Here we demonstrate radial velocity measurements and high-contrast imaging can be combined to overcome these issues. The presence of widely separated companions can be deduced by identifying an acceleration (long-term trend) in the radial velocity of a star. By obtaining high spatial resolution follow-up imaging observations, we rule out scenarios in which such accelerations are caused by stellar binary companions with high statistical confidence. We report results from an analysis of Doppler measurements of a sample of 111 M-dwarf stars with a median of 29 radial velocity observations over a median time baseline of 11.8 years. By targeting stars that exhibit a radial velocity acceleration (``trend'') with adaptive optics imaging, we determine that $6.5 \pm 3.0\%$ of M dwarf stars host one or more massive companions with $1 < m/M_J < 13$ and $0 < a < 20$ AU. These results are lower than analyses of the planet occurrence rate around higher mass stars. We find the giant planet occurrence rate is described by a double power law in stellar mass $M$ and metallicity $F \equiv $[Fe/H] such that $f(M,F) = 0.039^{+0.056}_{-0.028} M^{0.8^{+1.1}_{-0.9}} 10^{(3.8 \pm 1.2)F}$. Our results are consistent with gravitational microlensing measurements of the planet occurrence rate; this study represents the first model-independent comparison with microlensing observations.

\end{abstract}


\section{Introduction}
\label{S:Intro}

Over the past twenty years, numerous planets have been detected by several different techniques, permitting the first estimates of the occurrence rate of planets orbiting stars in the solar neighborhood \citep[e.g.][]{Johnson10b, Howard10b, Gould10, Vigan12}. As successful as these detection methods have been, each is sensitive only to a relatively narrow range of parameter space. For example, radial velocity (RV) studies are most sensitive to massive planets with orbital periods shorter than the time baseline of observations. 
\cite{Johnson10b} find that $3.4^{+2.2}_{-0.9}\%$ of M dwarfs have a Saturn-mass or larger planet within 2.5 AU. Beyond a few AU, RV searches are incomplete as the time required for a planet to complete one orbit is longer than the typical observing baseline. Some studies have attempted to extrapolate beyond this boundary. For instance, \cite{Cumming08} fit the observed RV planet population to a power law in planet mass and period and find that $18\% \pm 1\%$ of FGK stars host a Saturn-mass or larger planet within 20 AU. Recently, targeted RV surveys of M-dwarfs have suggested the giant planet occurrence rate is significantly smaller for these diminutive stars. \cite{Bonfils13} suggest fewer than $1\%$ of M-dwarfs host a Saturn-mass or larger planet with an orbital period $1 < P < 10$ days, and $2^{+3}_{-1}\%$ host giant planets with orbital periods between 10 and 100 days.

Transit studies suffer from similar detection biases. Since a planet transits only once each orbit, several orbits must be observed to definitively confirm a planet so characterization is limited to planets with periods shorter than a fraction of the observing baseline \citep{Gaudi05}. Additionally, the probability of a planet transiting its host star decreases with increasing orbital period \citep{Winn11}, such that hundreds of thousands of stars must be monitored in order to study the planet population at $a \approx 1$ AU \citep{Borucki84}. Nevertheless, the success of the \textit{Kepler} mission \citep{Borucki10, Koch10} has allowed for statistical analyses of transiting planets to be undertaken. For example, \cite{Morton13} analyze M dwarfs included in the 2012 list of announced Kepler Objects of Interest \citep[KOIs,][]{Batalha13}. By correcting for false positives (detections when no transiting planet exists), false negatives (nondetections when a transiting planet is present) and geometric effects (nondetections of nontransiting planets), they estimate an occurrence rate of $1.5$ planets with periods less than 90 days and radii larger than $0.5R_\oplus$ per M dwarf star. The occurrence rate found by these authors is slightly higher than previous analyses which measure rates of approximately one planet per star \citep{Youdin11, Mann12, Swift13, Dressing13}.

Neither RV nor transit searches are yet conducive to the discovery and characterization of planets well beyond the ``snow line,'' where water exists as ice. Instead, high contrast direct imaging techniques can be a powerful tool for detecting young planetary companions in this domain. The first direct imaging planet discoveries are securely in hand, including four companions to HR\,8799 \citep{Marois08, Marois10} and one each around $\beta$ Pictoris \citep{Lagrange09}, and Gl\,504 \citep{Kuzuhara13}\footnote{Companions detected around Fomalhaut \citep{Kalas08, Currie12}, HD 95086 \citep{Rameau13}, and LaCa15 \citep{Kraus12} are also good candidates to be directly imaged planets, but their true nature is somewhat ambiguous.}. Recent studies using these techniques have calculated an occurrence rate around A stars of $8.7^{+10.1}_{-2.8}\%$ at $1\sigma$ confidence for planets larger than $3 M_J$ and separations between 5 and $320 \textrm{ AU}$ \citep{Vigan12}. Imaging studies have been most effective around high mass stars. \citep{Crepp11, Carson13}. Nondetections around lower-mass stars have been used to place upper limits on the frequency of giant planets. For example, \cite{Nielsen10} rule out the presence of giant planets orbiting FGKM stars beyond 65 AU with $95\%$ confidence. High contrast imaging, while powerful, only provides a measure of the relative brightness of a companion. To estimate the companion's mass, the age of the star must be known and planetary thermal evolution models must be applied to estimate the temperature (and brightness) of the companion \citep{Chabrier00, Baraffe03}. Moreover, direct imaging is currently only sensitive to massive planets; the HR8799 planets and $\beta$ Pic b are believed to have masses $m > 5 M_J$. RV and transit studies suggest such ``super-Jupiters'' are rare compared to Jovian-mass and smaller objects at smaller separations \citep{Howard10b, Howard12}.

The gravitational microlensing technique is also effective for finding giant planets in wide orbits and does not rely on planetary evolution models. Using this technique, planets can be detected by observing perturbations to the photometric gravitational microlensing signal when a planet and its host pass in front of a more distant star. Since $70-75\%$ of stars in the galaxy are M-dwarfs, most lenses have mass $M < 0.5 M_\odot$. Microlensing searches thus provide a measure of planet occurrence around low mass stars. Microlensing studies are sensitive to planets near the Einstein ring, $R_E \sim 3.5 \mbox{AU} (M/M_\odot)^{1/2}$, a much wider separation than RV and transit searches \cite{Gould10}. \cite{Cassan12} find microlensing searches are most sensitive to planets at a projected separation in the range $[s_{\textrm{max}}^{-1}R_E, s_{\textrm{max}}R_E]$, where $s_{\textrm{max}} \sim (q/10^{-4.3})^{1/3}$ and $q$ is the mass ratio between a companion and the host star. These authors find a planet occurrence rate that can be parameterized by a double power-law function, in mass ratio $q$ and separation $s$, such that
\begin{equation}
\frac{d^2N}{d\log q d \log s} = 10^{-0.62 \pm 0.22} \bigg(\frac{q}{5 \times 10^{-4}}\bigg)^{-0.73 \pm 0.17} \mbox{dex}^{-2}.
\label{microlensingresults}
\end{equation}
The normalization constant is equivilent to $0.24^{+0.16}_{-0.10}$. These results are calculated under the assumption that planets are distributed uniformly in $\log s$, as is the case for binary stars \citep{Opik24}. Additionally, \cite{Sumi10} find a power-law slope in mass such that $dN/ d\log q \propto q^{-0.68 \pm 0.20}$ for Neptune-sized planets, but do not attempt to quantify a normalization factor. 

As microlensing studies focus on distant M-dwarfs ($d > 1$ kpc) in the direction of the galactic bulge \citep{Gaudi02}, these stars can be difficult to characterize accurately due to crowding. Stellar masses and metallicities are often estimated without being measured spectroscopically. If these host stars have different masses than assumed, it would affect the results of planet occurrence rate studies by microlensing groups as microlensing results do not account for correlations between stellar and planet properties. Additionally, as microlensing searches are most sensitive near $r = R_E$, beyond approximately 10 AU the lensing signal becomes very weak, and differentiating distant planets from unbound, ``free-floating'' planets becomes difficult \citep{Sumi11}.

RV and microlensing studies probe different regions around a star, and extrapolations between the two domains suggest a possible discrepancy. \cite{Cassan12} estimate a total giant planet occurrence rate significantly lower than the \cite{Cumming08} RV result.
Derived power-law distributions in mass may also be different for planets found by each method: \cite{Cumming08} find a distribution such that $dN/d \log m \propto m^{-0.31 \pm 0.20}$ from RV-detected planets, while \cite{Cassan12} find a distribution such that $dN/ d\log q \propto q^{-0.73 \pm 0.17}$. Since microlensing studies target M-dwarfs, which are confined to a narrow mass range, we can approximate $q = m/M$ as $m$. In this case, the microlensing result and RV result differ by $1.6 \sigma$.  Since giant planet occurrence decreases with decreasing stellar mass and metallicity \citep{Johnson10a}, the expected giant planet occurrence rate around M dwarfs would be smaller than that for FGK stars. Therefore, it is necessary to compare the microlensing planet population not to a population of FGK stars, but instead to a study of RV detected planets around M-dwarfs.

Historically, RV observations have been used to detect and characterize planets once they complete a full orbit, limiting studies to planets with periods shorter than the observing time baseline. In this paradigm, potentially useful information is overlooked. Wide companions are not completely undetectable: instead they can be identified by the presence of long-term RV accelerations (linear ``trends'')  which can be used to infer the existence of a companion in a more distant orbit \citep{Liu02, Crepp12a}. However, a linear acceleration does not provide unique information about the mass and period of the companion: the same trend could be caused by a Jupiter-mass planet at 5 AU or a $100 M_J$ M-dwarf at 25 AU. This degeneracy can be broken by adaptive optics (AO) imaging. Low-mass binary companions to nearby M-dwarfs can be easily imaged by modern AO systems \citep{Lloyd02, Siegler03}. Such detections form the basis for the TRENDS High-Contrast Imaging Survey, which to date has detected four M-dwarfs and one white dwarf companion to higher mass stars\citep{Crepp12b, Crepp13a, Crepp13b}.

In this work, we combine RV and AO observations of nearby cool stars to estimate the frequency of giant planets in wide orbits around M-dwarfs. From a sample of 111 M-dwarfs observed with a median Doppler RV baseline of 11.8 years, we identify 4 systems with long-term RV accelerations but no known companions and target these stars with AO imaging in an attempt to detect stellar-mass companions. We discuss these observations and our methodology in \textsection\ref{SO}. Given an observed RV trend or lack thereof, we determine with high statistical confidence if a giant planet exists around each star. We analyze the effects of false positive and false negative detections of RV accelerations in our sample in \textsection\ref{Stats}. In \textsection\ref{Results} we estimate the occurrence rate of giant planets around M-dwarfs and compare the measure to results from other techniques. We summarize and conclude in \textsection\ref{SC}.

 This study represents the first measurement of the planet population in the range 0--20 AU. While we rely on brown dwarf cooling models, our study does not make use of theoretical planetary evolution models, unlike other AO studies of planetary systems.

\section{Sample and Observations}
\label{SO}
\subsection{Target Selection}
\label{TS}

Since 1997, the California Planet Search (CPS) collaboration has undertaken a comprehensive Doppler search for extrasolar planets at the Keck Observatory \citep[e.g. ][]{Howard10}. Using Keck/HIRES \citep{Vogt94}, the CPS program monitors over 2000 stars, most selected to be chromospherically quiet, single, and bright. Included in this sample is a collection of M-dwarfs from the Gliese and Hipparcos catalogs brighter than $V = 11.5$ and lacking known stellar companions within 2 arcseconds \citep{Rauscher06}. This sample was later extended to $V = 13.5$ and currently includes 131 M-dwarfs within 16 pc of the Sun, where we define the M spectral class as targets with $B-V > 1.44$.

To develop the sample used here, we first remove from this set 16 stars with a known, nearby stellar binary companion. We define ``nearby'' as a separation small enough that a test particle orbit with semimajor axis $\geq 30$ AU would be unstable, following the instability criterion of \cite{Holman99}. This criterion depends on the unknown eccentricity of the binary pair, as perturbative effects are maximized at periapsis. We take $e=0.5$ as a typical value and find the onset of instability occurs for binary stars with $a \sim 250$ AU. Planets can still form in these more compact binary systems \citep[e.g. Gl667C; ][]{Anglada-Escude12b} but at such small separations protoplanetary disk formation and planet evolution would be affected significantly by the presence of stellar companions. This selection thus allows us to study a class of planets that likely followed similar evolutionary processes. Moreover, the detection of an acceleration around these stars is ambiguous, as it could be caused by the binary star, a planetary-mass companion, or both together.

\begin{figure*}[htbp]
\centerline{\includegraphics[width=1.0\textwidth]{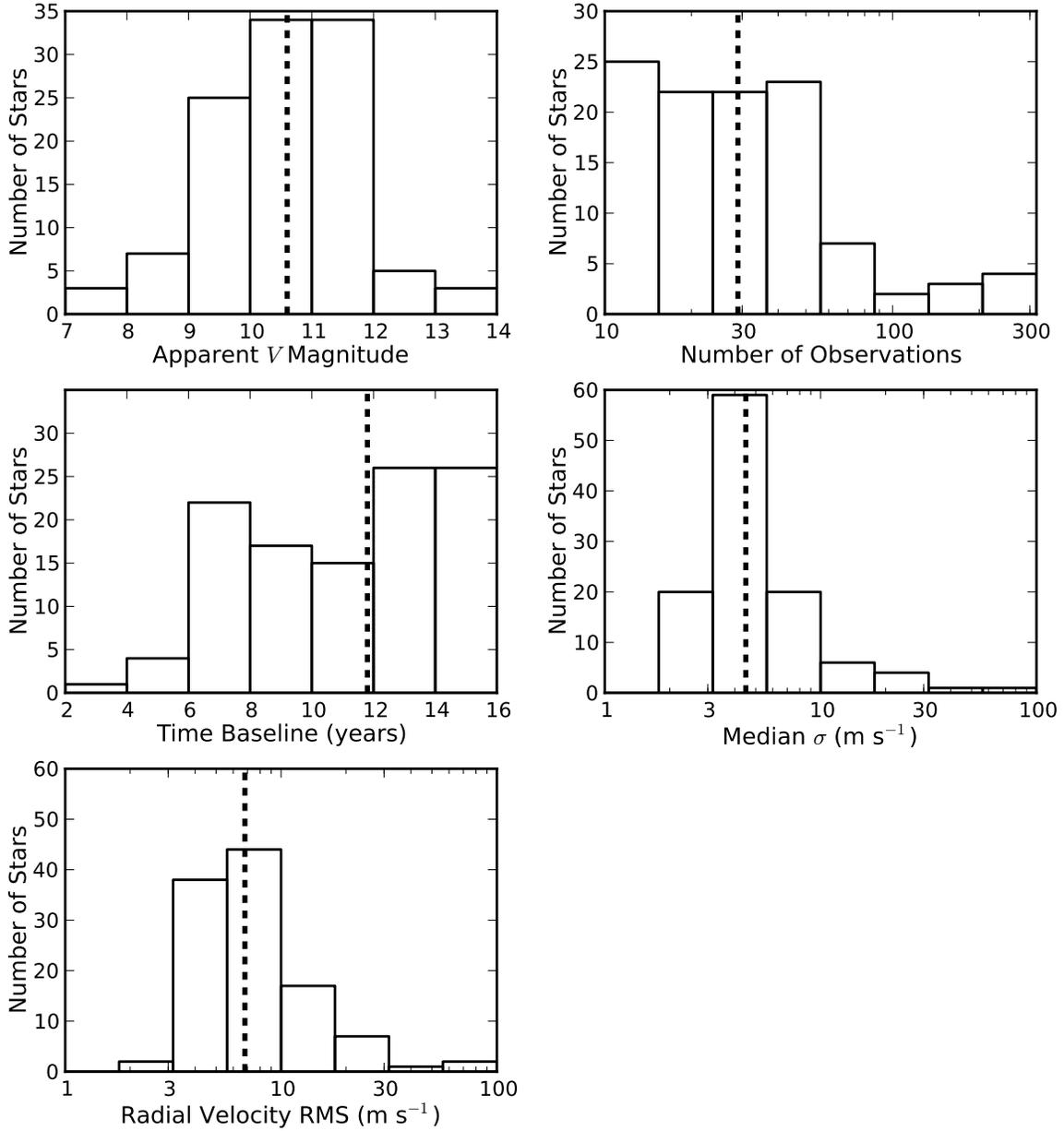}}
\caption{Distributions of the RV observational parameters. Dashed lines represent the median values for each parameter. The median target brightness is $V = 10.6$, and the median target has been observed 29 times over 11.8 years. The median measurement uncertainty $\sigma$, defined as the sum in quadrature of rotational jitter and statistical uncertainty (Eq. \ref{sigmaeq}) is 4.5 m s$^{-1}$. Specific parameters for each individual system are shown in Table \ref{T1}.
  }
\label{Hists}
\end{figure*}

\begin{figure}[htbp]
\centerline{\includegraphics[width=0.5\textwidth]{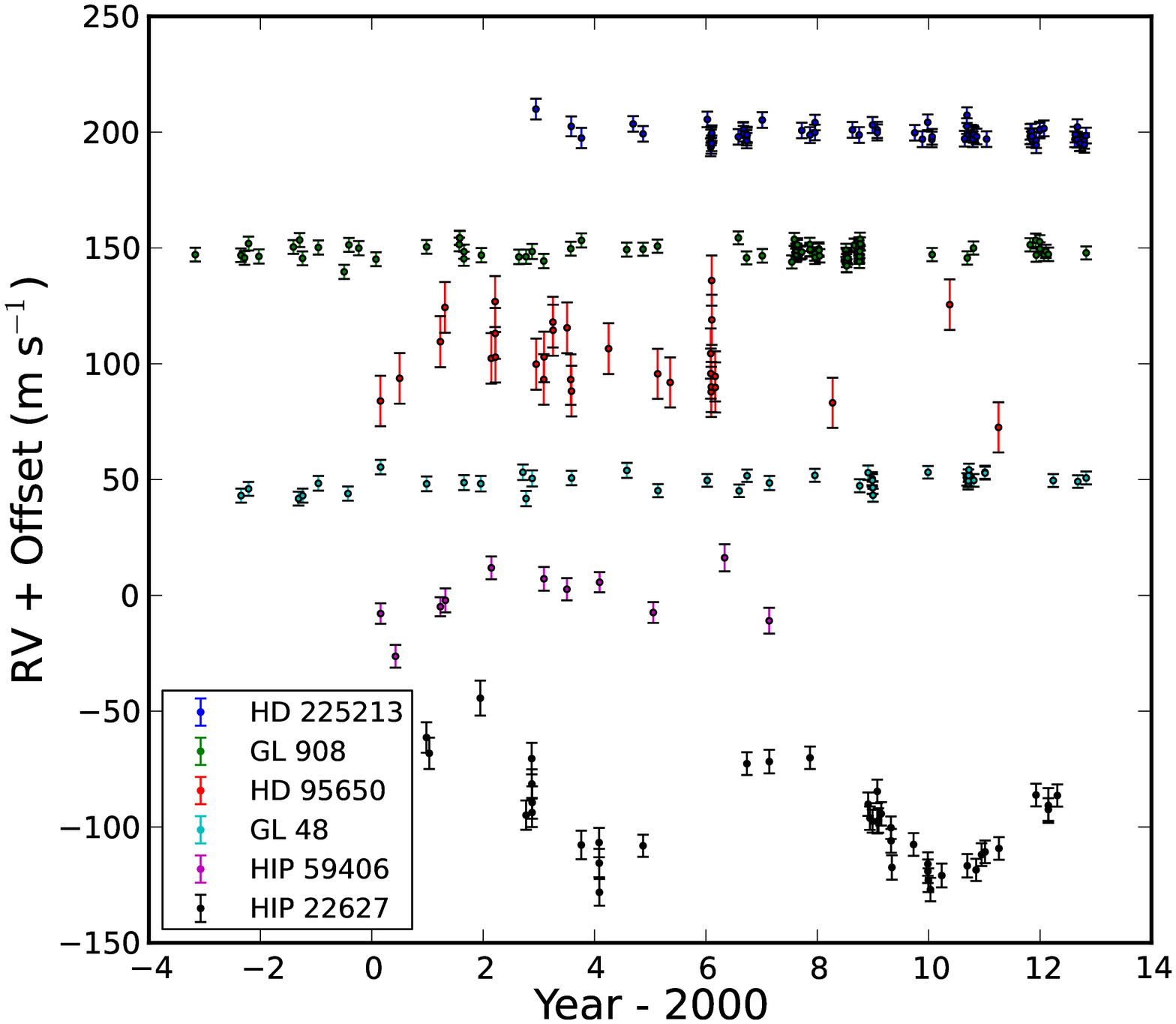}}
\caption{RV measurements for a representative sample of six example stars. The stars are arranged such that the brightest star is at the top of the plot. The individual stars vary considerably with respect to observing baselines, measurement uncertainty, and number of observations. Of these stars, HIP\,59406 has a wide binary companion, while HIP\,22627 has both a known inner planet and long-term RV acceleration.
  }
\label{RVfig}
\end{figure}

After making the above selection we are left with 111 RV targets, all of which have at least 8 radial velocity observations and a time baseline longer than 2.9 years. The median number of observations is 29 over a median time baseline of 11.8 years. The stars have spectral types from M0 to M5.5 and masses in the range $0.64M_\odot - 0.10M_\odot$. Stellar masses are estimated using the empirical relation between mass and absolute K-band magnitude, $M_K$, described by \cite{Delfosse00}. We take $10\%$ as a typical uncertainty in the stellar mass, in line with previous estimates \citep{Bean06}. $K$-band apparent magnitudes are measured using apparent magnitudes from the 2MASS point-source catalog \citep{Cutri03}. The majority of our parallaxes are taken from an analysis of \textit{Hipparcos} data \citep{VanLeeuwen07}. Some of our stars were not observed by Hipparcos, while others have had their distances updated more recently. In these cases, we apply the distances listed in the SIMBAD astronomical database (Table \ref{T1}). For example, for Gl\,317 we use the parallax found by \cite{Anglada-Escude12a}; their derived mass and metallicity are consistent with our estimated values. In all cases, stellar metallicities are estimated by measuring the offset between the star's position in the \{$V-K_s, M_{K_S}$\} plane from a calibrated main sequence following the method of \cite{Neves12}. We take 0.17 dex as a typical uncertainty in the stellar metallicity, representative of the scatter between this photometric method and spectroscopic measures of stellar metallicity. Stellar parameters for these targets are listed in Table \ref{T1} and observational parameters are listed in Table \ref{T1b}. The distribution of RV observational parameters are shown in Fig. \ref{Hists}. Spectral types are estimated by comparing the spectrum collected with HIRES to other spectra collected with this same instrument. RV observations for a representative sample of six ``typical'' stars are shown in Fig. \ref{RVfig}. 

\subsection{Detecting Accelerations from Radial Velocities}
\label{RVs}

\begin{figure*}[htbp]
        \centering      
\centerline{\includegraphics[width=\textwidth]{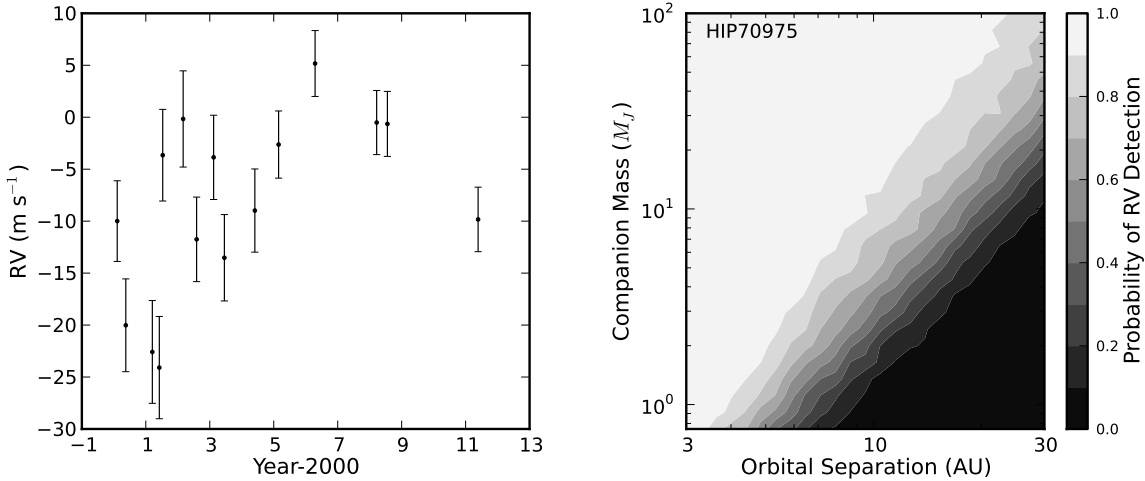}}
\caption{(left) RVs for HIP\,70975, a typical star in our survey. This $0.32 M_\odot$ M-dwarf has a total of 15 radial velocity observations over a baseline of 15.5 years, with an average RV precision (including photon noise and jitter) of 4 m/s. (right) Detectability plot showing the likelihood of an RV detection for a companion orbiting HIP\,70975 as a function of companion mass and semimajor axis from its host star. \newline
  }
\label{Detectability}
\end{figure*}

The detection of a long-term RV acceleration is facilitated by having many observations over a long time baseline to increase signal, but complicated by astrophysical ``jitter'' caused by rotational modulation of surface inhomogeneties. To determine the masses and semimajor axes to which we are sensitive to planetary companions, we inject a series of artificial companions into orbit around the stars in our sample. We define a logarithmically spaced grid of companion masses and semimajor axes spanning the range $0.75 M_J < m < 100 M_J$ and $3 \textrm{AU} < a < 30 \textrm{AU}$, such as the one shown in Fig. \ref{Detectability}. At each point, we inject 500 planets and randomly assign each of the remaining orbital elements. The longitude of ascending node $\Omega$, time of periapsis $t_p$, and argument of periapsis $\omega$ are drawn from a uniform distribution, while the inclination is drawn from a distribution $dn/di = \sin i$ and the eccentricity from a distribution such that $dn/de$ follows a beta distribution with $a = 1.12$ and $b = 3.09$, which well-replicates the distribution of observed eccentricities for RV planets with orbits longer than 382 days \citep{Kipping13}. We then numerically integrate these orbits forward in time over our true observing baseline.

At the epochs each star was observed by CPS, we calculate the expected radial velocity signal caused by our injected planet. Each velocity is perturbed from the true expected Keplerian velocity by a normal variate with zero mean and standard deviation $\sigma$ representative of the total expected noise:
\begin{equation}
\sigma = \sqrt{\sigma_\gamma^2 + \sigma^2_\textrm{jitter}}.
\label{sigmaeq}
\end{equation}
Here, $\sigma_\gamma$ is the photon noise, estimated for each individual observation by randomly selecting a single measurement of the measured Poisson photon noise from a true observation of the star. To account for the effects of jitter, we follow the method of \cite{Isaacson10}, who develop an empirical relation between the level of stellar jitter, a star's $S_{\rm HK}$ value, and its $B-V$ color. $S_{\rm HK}$ is defined as the ratio of the flux in the Ca II line cores to flux in the surrounding continuum. We compare the $S_{HK}$ value observed by CPS to that expected from the star's $B-V$ color, which provides an estimate of $\sigma_\textrm{jitter}$. This value is added in quadrature to the photon noise to estimate a total observational uncertainty, $\sigma$. Typical observations carry a photon noise of $2-4$ m s$^{-1}$ and jitter values are typically $3-5$ m s$^{-1}$ for a total $\sigma$ value of $3-6$ m s$^{-1}$ for the majority of stars. Median $\sigma$ values for each star are listed in Table 2.

Once all observations are accounted for, we search for evidence of our injected planetary companion, manifested as an acceleration in the RV data. Here, we define the existence of a trend using the Bayesian Information Criterion \citep[BIC;][]{Schwarz78, Bowler10, Campo11, Stevenson12}, which prefers simple, well-fitting models subject to 
\begin{equation}
\rm{BIC} \equiv -2 \ln \mathcal{L} + k \ln N,
\label{EqBIC}
\end{equation}  
where $\mathcal{L}$ is the maximum likelihood for a model with $k$ free parameters and $N$ observations. The BIC thus favors models that fit the underlying data well, but penalizes increasingly complex models. For a more complex model to be preferred by the BIC, it must improve the fit by an amount greater than $k \ln N$ to overcome the penalty term.

\cite{Kass95} claim a difference between BIC values provides a bounded approximation of twice the logarithm of the Bayes factor. A change in BIC value of ten or more (corresponding to a Bayes factor of approximately 0.01) suggests strong evidence for an association between two parameters. If the BIC value decreases by more than 10 when considering a model with a linear acceleration over a model with only an offset, a planet is considered to be detected. Otherwise, the system is considered a non-detection. We find that the $\Delta$BIC value chosen here is consistent with by-eye inspection of our data in a visual search for RV accelerations. In both cases, we allow for a linear offset in the RV data in August, 2004, corresponding to an upgrade of the HIRES CCD detector \citep{Wright11}. Effectively, we treat the data from before and after the upgrade as coming from two distinct instruments, which serves to slightly decrease our sensitivity to small RV accelerations.

\begin{figure}[htbp]
        \centering      
\centerline{\includegraphics[width=0.5\textwidth]{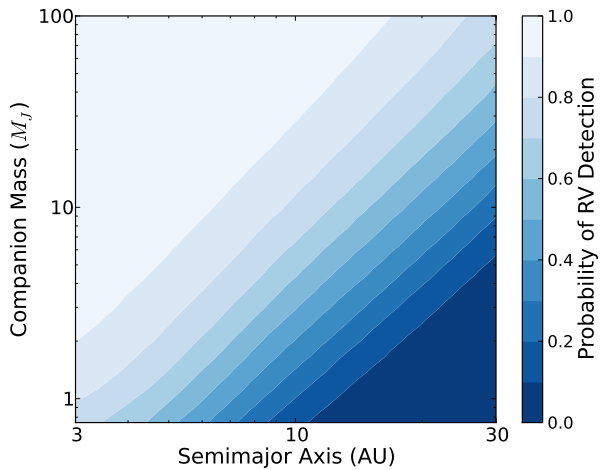}}
\caption{The ensemble average likelihood over all 111 stars of an RV detection for a companion to a star in our sample as a function of companion mass and orbital semimajor axis. We can detect accelerations induced by planets as small as $1 M_J$ in short orbits, but a planet distribution function is required to determine the number of $1 M_J$ planets in wide orbits and calculate the overall giant planet occurrence rate.
  }
\label{DetectFull}
\end{figure}

By repeating this process for many simulated planets over our mass-semimajor axis grid, we can map out the relative probability of detecting a linear trend caused by a planet as a function of companion mass and semimajor axis. As an example, Fig. \ref{Detectability} shows RVs for HIP\,70975 and the likelihood of detecting a planet at a given mass and period given these observations. Fig. \ref{DetectFull} shows the mean likelihood of detecting a planet around a given star across our sample. Throughout this work, we report the occurrence rate of planets with masses in the range $1 M_J < m < 13 M_J$. We can detect accelerations caused by planets smaller than $1 M_J$ in certain instances, but would miss the majority of these planets. As Fig. \ref{DetectFull} shows, we can only detect a $0.75 M_J$ planet at 6 AU $50\%$ of the time; planets at smaller separations would exhibit significant curvature over a 12 year time baseline and could be detected through an RV survey alone. We are more efficient at detecting planets larger than $1 M_J$, although we would still not expect to detect all planets in this range. We account for false negative ``missed'' planets in our analysis, as described in \textsection\ref{FN}.

Eight of the stars in our sample host known planets with closed orbits. All of the planets have $m \sin i < 2.5 M_J$ and are listed in Table \ref{knownpl}. To identify radial velocity accelerations caused by outer planets, we include the signal from these planets by comparing a model which contains the known planet and an acceleration to a model which contains only the known planet. Two known planets in our sample, Gl 876b and Gl 317b, are larger than $1 M_J$, so in addition to searching these systems for long-term RV accelerations, we also include these known planets in our giant planet occurrence calculations.

One additional planet, Gl\,649b, has a best-fitting mass $m \sin i = 0.90 \pm 0.05 M_J$; if the inclination is smaller than 64 degrees this planet has mass $ m > 1 M_J$. We follow the method of \cite{Ho11} to determine the probability of this event. That is, we define the probability that the true mass $m$ is greater than some value $X$ given an observed mass $m_O = m \sin i$ such that
\begin{equation}
P(m > X | m_O) = 1 - \frac{\int_{m_O}^X \frac{(m_O/m^2)}
{\sqrt{1-(m_O/m)^2}}P(m)dm}{\int_{m_O}^{m_\textrm{max}}
\frac{(m_O/m^2)}{\sqrt{1-(m_O/m)^2}}P(m)dm}
\end{equation}
Here, $P(m)$ is the true planet mass distribution function.
$m_\textrm{max}$ is the physical upper mass limit for a planet. Since the true distribution function is strongly biased towards small planets, the number selected here does not significantly affect our results. By simply assuming the star is aligned randomly along our line of sight so that the inclination distribution is flat in $\cos i$, the result of a flat planet mass distribution function, we expect a observed mass $m \sin i = 0.90 M_J$ to be produced by a Jupiter-mass or larger planet 56\% of the time; all reasonable assumptions of an underlying mass distribution affect this value by less than 10\%. We repeat this procedure for all confirmed planets in our sample with masses $m \sin i < 1 M_J$ to quantify the likelihood that other known planets are $m > 1 M_J$ planets with low inclinations. We find, in addition to Gl\,849b, HIP\,22627b ($m \sin i = 0.64 M_J$) has approximately a 25\% probability of having a mass $m > 1 M_J$. This probability is vanishingly small for all other known planets.

\begin{figure*}[htbp]
\centerline{\includegraphics[width=0.95\textwidth,trim={0 0 0 0}, clip=true]{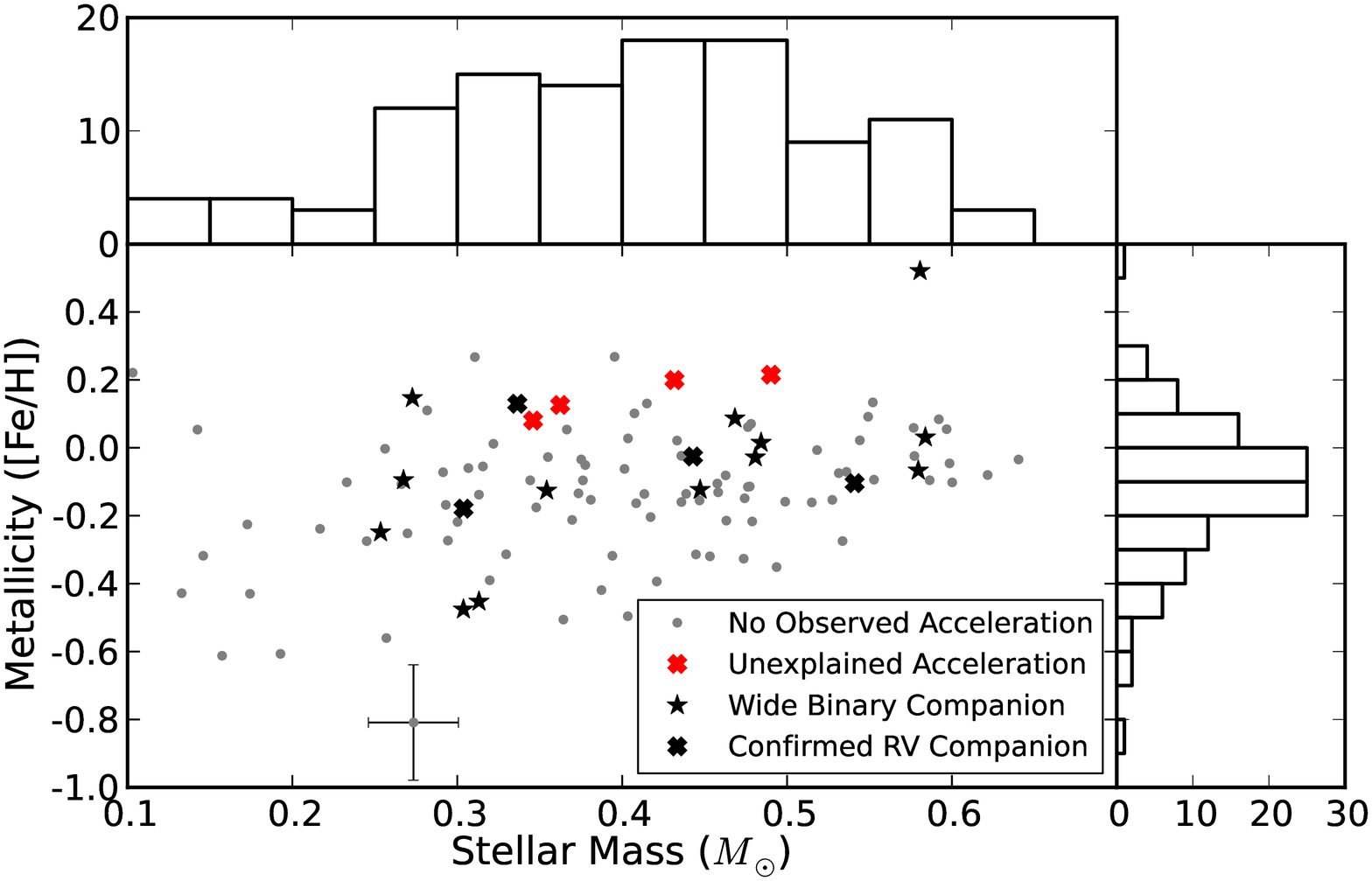}}
\caption{Observed M-dwarf sample in the stellar mass-metallicity plane. Systems with observed RV accelerations are shown in red while those without a detected acceleration are in black. Systems with a wide binary companion are labeled with stars, while diamonds represent systems with confirmed planets of any mass. The error bars displayed for HD\,33793 are representative of the uncertainties for all stars in our sample.
  }
\label{massmetal}
\end{figure*}

Of our sample of 111 stars, 2 have confirmed planets larger than $1 M_J$, 6 systems have confirmed RV planets with masses $m \sin i < 1 M_J$ only, two exhibit RV acceleration caused by known brown dwarfs, and four show unexplained long-term RV accelerations, such that $\Delta$BIC > 10 when we include an acceleration term in our fit to the RV data. In the case of Gl\,849b, the long-term acceleration exhibits significant curvature, so we are able to place constraints on this object's mass and orbital semimajor axis (see Appendix). In all other cases, the magnitude of the observed acceleration is different from zero by $3\sigma$. Additionally, the magnitude of the acceleration is such that over the observing baseline, the expected $\Delta\textrm{RV}$ induced by the putative outer planet is larger than the uncertanties of each individual data point. The distribution of these systems in the stellar mass-metallicity plane is shown in Fig. \ref{massmetal}.

For the four targets with an observable RV drift, we create a grid of logarithmically-spaced companion masses and semimajor axes over the range $0.75 < m/M_J < 100$ and $3 \textrm{AU} < a < 30 \textrm{AU}$. For a given grid point, we determine the best-fitting Keplerian orbit for a given eccentricity and inclination. We assume the inclination and eccentricity distributions are the same as assumed previously. The eccentricity distribution is well-characterized for solar-type stars, but may not hold for planets around lower-mass stars. We find the exact choice of eccentricity distribution does not significantly affect our results.

\begin{figure*}[htbp]
\centerline{\includegraphics[width=1.0\textwidth, trim={0 0 0 0}, clip=true ]{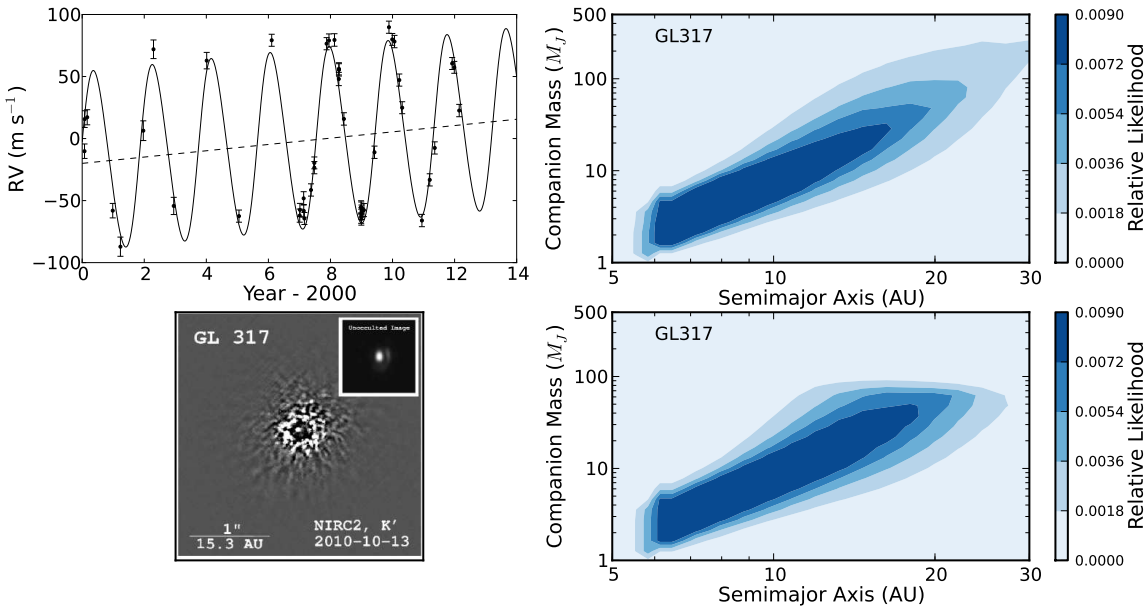}}
\caption{(top left) RV observations for Gl\,317 over our 12.1 year baseline. The best-fitting RV acceleration is $-2.51 \pm 0.62$ m s$^{-1}$ yr$^{-1}$ (dashed line); the best-fitting model which includes both the planet and the acceleration is shown as a solid line. (top right) Probability contours marginalized over eccentricity and inclination, displaying the location of a giant companion orbiting Gl\,317 from RVs alone. The likelihood values are normalized such that the sum of the likelihood over our 26x25 grid of companion masses and separations sums to unity. (bottom left) AO image of Gl\,317, showing no companion is visible in the AO imagery, either in the unocculted image (inset) or when a coronagraph is inserted. This eliminates the possibility of a stellar-mass companion at a projected separation smaller than 48 AU. (bottom right) Probability contours displaying the location of a giant companion to Gl\,317 when the RV data is combined with AO data. We find the RV acceleration is likely induced by a substellar companion.
  }
\label{WrightRV}
\end{figure*}

We determine the likelihood of the best-fitting orbit for each mass, period, eccentricity, and inclination. We then convert these likelihoods into relative probabilities, assuming our errors are uncorrelated so that $P \propto(-\exp(\chi^2/2)))$. We then marginalize over eccentricity and inclination and normalize our probabilities so that $\displaystyle\sum\limits_{M,a} P = 1$. In these cases, we assume the inclination is random on the sky, so that the inclination follows the distribution $f(i) = \sin i$. Assuming a different planet mass distribution function affects this result by less than 10\%. The result is a contour in the mass-semimajor axis plane for the likelihood that a given object could cause the observed stellar radial velocity variation \citep{Wright07}. An example is shown in Fig. \ref{WrightRV}. Implicit in this analysis is the assumption the radial velocity variation is dominated by the motion of a single, massive companion rather than the constructive interference of the RV signal of two or more smaller objects. We discuss false positive probabilities in \textsection\ref{FP} and conclude the assumption that one signal dominates the observed RVs is reasonable.

The magnitude of an acceleration depends on both the semimajor axis and mass of the companion. For a planet in a circular orbit, the magnitude of the change in radial velocity, $\dot{\gamma} = dv/dt$, is given by
\begin{equation}
\dot{\gamma} = (6.57 \mbox{ m} \: \mbox{s}^{-1} \mbox{yr}^{-1})\bigg(\frac{m_p}{M_J}\bigg)\bigg(\frac{a}{5 \mbox{AU}}\bigg)^{-2} \hat{v}_p \cdot \hat{r}_{los},
\label{TrendSize}
\end{equation}
with $M_J$ the mass of Jupiter and $a$ the orbital semimajor axis. $\hat{v}_p$ and $\hat{r}_{los}$ are unit vectors along the direction of the planet's velocity vector and the line of sight, respectively. When the companion has longitude of periapsis $\varpi = 90$ or $270$, the magnitude of this trend is maximized: $\hat{v}_p \cdot \hat{r}_{los} = \sin i$. To determine if our observed accelerations are caused by planets or more massive companions, we obtained AO imaging observations of each star.

\subsection{Adaptive Optics Observations}
\label{AO Description}
The detectability diagnostics developed in \textsection\ref{RVs} are based strictly on the information encoded in the RV data. Since we are looking at accelerations caused by objects in wide orbits around the primary star, we must break the degeneracy between companion mass and orbital semimajor axis for a given observed acceleration. AO imaging allows us to immediately detect the presence or nonexistence of nearly all stellar-mass companions and most brown dwarf companions to our primary stars, so we can readily separate stellar-induced accelerations from those caused by planets. 

All four targets with an observable RV acceleration were observed with NIRC2 (instrument PI: Keith Matthews) at the W.M. Keck Observatory using the AO system \citep{Wizinowich00} (Table \ref{T2}). In most cases, images were obtained in the $K^\prime$ filter ($\lambda_c = 2.12 \mu$m). We nominally execute a three-point dither pattern to facilitate removal of instrument and sky background noise. Images were processed by flat-fielding, correcting for hot pixels with interpolation, subtracting the sky background, and rotating the frames to standard north-east orientation. In three cases, we applied the angular differential imaging (ADI) point spread function subtraction technique, allowing the observed field to rotate around the target star during the observation, while instrumental artifacts remain fixed. In all cases, we use the large hexagonal pupil mask and the narrow camera. For all four systems exhibiting long-term RV accelerations, we did not image a massive companion. In the cases where our field of view is not large enough to eliminate the possibility of massive stars in very wide orbits ($ > 4''$), we supplement our AO data with publicly available 2MASS images.

\begin{figure}[htbp]
\centerline{\includegraphics[width=0.5\textwidth]{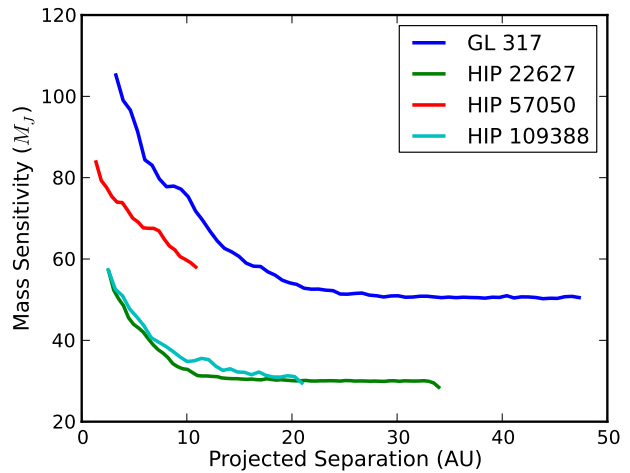}}
\caption{Mass sensitivity for a $5\sigma$ detection of a companion object as a function of projected angular separation for each of our four stars with long-term RV drifts. The maximum projected separation eliminated corresponds to the field of view of the AO system and thus varies for each star as a function of the distance to each star. For all stars except HIP\,57050, we rule out stellar mass companions beyond 1 arcsecond through our adaptive optics imaging. When our field of view is small, we supplement our AO data with 2MASS seeing-limited images. Stellar companions at small projected separations would have RV accelerations larger than those observed in our sample. 
  }
\label{CCplot}
\end{figure}

The luminosity ratio between our M-dwarfs and their companions depends on the mass of the companion and the age of the system. Stars observed by the CPS team are selected to avoid excessive chromospheric activity, and are thus likely older than 1 Gyr \citep{Wright05}. We assume all targets have fully contracted and assert an age of 5 Gyr for each system. For systems with nondetections, we estimate the flux (and thus the mass) a companion would need to have to be observed at a given projected separation in our observations. From that value, we can then determine the region of parameter space excluded by the observations (Fig. \ref{CCplot}). In general, AO imaging eliminates nearly all stellar companions, while ADI can also probe the brown dwarf mass regime.

For each of our targets with unexplained accelerations, a contrast curve showing the mass to which we are sensitive to companions at the $5\sigma$ level as a function of projected separation is shown in Fig. \ref{CCplot}. This choice provides similar results to the detection limits found by visual inspection, as tested by injecting artificial companions into AO images \citep{Metchev09}. We convert relative brightness to mass using the theoretical evolutionary tracks of \cite{Baraffe03} for substellar companions and \cite{Girardi02} for more massive companions. Interpolation between the two sets of models provides reasonable results in the intermediate domain near $125 M_J$. The resultant parameter space where a companion could reside to cause the observed stellar acceleration is shown in Fig. \ref{WrightRV}.

\begin{figure}[htbp]
\centerline{\includegraphics[width=0.5\textwidth]{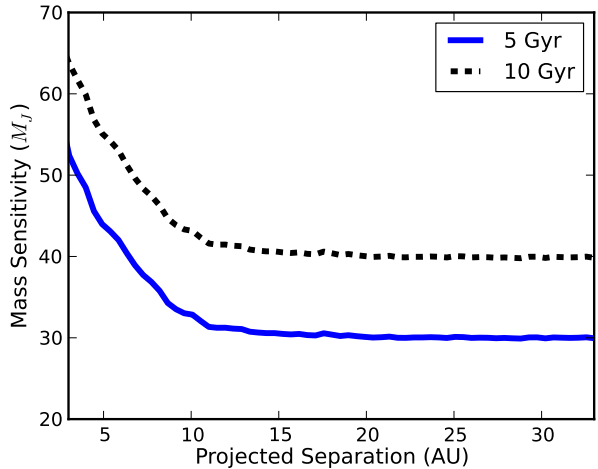}}
\caption{Adaptive optics mass exclusion plot for the star HIP\,22627 showing the relative insensitivity of our results to the assumed age of M-star planet hosts. Adaptive optics observations rule out essentially all stellar mass companions. Sensitivity to substellar objects is a function of age, but brown dwarfs are scarce at close separations \citep{Marcy00}, and wide separations \citep{Metchev09}. Thus, our estimate of the planet frequency around M-stars is only weakly dependent on our assumed age of the host stars.  
  }
\label{AgeAO}
\end{figure}

The assumption of a 5 Gyr age for each star does not significantly affect our results. For all plausible system ages, stellar mass companions would be easily detectable by AO. Our sensitivity to stars is independent of assumed age, as luminosities of M-dwarfs are constant over the age of the universe. At no ages $> 1$ Gyr are we sensitive to any planetary mass companions. As shown in Fig. \ref{AgeAO}, assuming a different age for each star would only change the efficiency of detecting brown dwarfs. Since the occurrence rate of brown dwarfs is only a few percent, much smaller than the occurrence rate of planets or low-mass stars \citep{Metchev09, Dieterich12}, errors induced by assuming an incorrect stellar age from missed brown dwarfs are small. ``False negatives'' such as these will be discussed in \textsection\ref{FN}.

\section{Measuring the Giant Planet Occurrence Rate}
\label{Stats}

We estimate the occurrence rate of giant planets orbiting M-dwarfs using statistical inference. The fraction of stars which host giant planets, given some number of observed accelerations $N\scriptsize{\mbox{trends}}$ and some number of nondetections $N\scriptsize{\mbox{ND}}$ from a sample of targets, is given such that
\begin{equation}f_{pl} = \frac{N\scriptsize{\mbox{trends}}\normalsize P(\mbox{planet}|\mbox{trend}) + N\scriptsize{\mbox{ND}}\normalsize P(\mbox{planet}|\mbox{ND})}{N\scriptsize{\mbox{targets}}}
\label{ResultEq}
\end{equation}
To calculate the posterior probability that a given star hosts a gas giant planet, we must estimate the \textit{a priori} likelihood that a planet exists given the presence of an RV acceleration (a true positive), the likelihood a planet would not be detected in an RV survey (a false negative) and the likelihood that an observed acceleration is caused by some effect other than the movement of a planet (a false positive). 

\subsection{False Negatives}
\label{FN}

There are multiple ways for a giant planet to be missed in our survey. For each planet in a wide orbit, we observe only a fraction of a revolution. A planet near its maximal sky-projected separation from its host star has acceleration primarily in the tangential, not radial, direction. In cases such as this the change in radial velocity over our observing baseline may not be noticeable. Thus we may expect to have a lower RV detection efficiency for planets near their maximal sky-projected separation. 

Similarly, we may expect to have a lower imaging detection efficiency for stars near their minimal sky-projected separation, when the RV acceleration is the largest. However, in these cases we would still expect to detect the binary companion. If the companion is located directly along the line of sight to the star, then it will also appear in the 0''.85 spectrograph slit used with HIRES. Therefore, we would expect such systems to appear as SB2s. We explore this fully, and show that we would detect all such systems, in \textsection\ref{PMBS}.

To determine the likelihood that such a planet would be missed by our search, we use our detectability matrices developed in \textsection\ref{RVs}. We assume the distribution of planets follows a double power-law, such that
\begin{equation}
\frac{d^2N}{d\log m d\log a} \propto m^{\alpha} a^{\beta},
\label{DPL}
\end{equation}
similar to that assumed by \cite{Cumming08} and \cite{Bowler10}, and comparable to the power-law distributions applied in the analyses of microlensing surveys. At a given companion mass and semimajor axis, we can then determine the relative likelihood that a planet exists at this position. We multiply this by the likelihood of detecting such a planet to determine the fraction of planets we would find orbiting each star and the fraction we would miss. These numbers are determined through our analysis of observations of simulated injected planets, as developed in \textsection \ref{TS}.

We can test our detectability calculation by analyzing the known wide-separation companions in our sample. Of our 111 stars, four are known to host directly imaged brown dwarf companions (see Appendix). Of these, two (HD\,71898B and HIP\,63510B) were detected as accelerations in our sample, while two (Gl\,569B and Gl\,229B) are at very large separations and were not detected. The detection or nondetection of each system is consistent with what would be expected from our analysis of injected planets (see Appendix).

We detect the two brown dwarfs with high expected RV detection efficiency, and do not detect the two with expected detection efficiencies near zero, both of which have $a > 40$ AU. We would like a larger sample to test this method, but the limited number of brown dwarfs suggests our ability to detect giant planets is consistent with expectations. This sample also suggests $f_{BD}$ is only a few percent, consistent with complementary studies \citep{Dieterich12}.

A giant planet could also be missed if it was in a system with multiple giant planets. We observe only the sum of all radial velocity signals from all planets orbiting a star. For example, if a star hosts two giant long-period planets with one on each side of the star, the two signals would destructively interfere. Even if the acceleration was still detectable, this interference would cause us to measure an incorrect magnitude of the acceleration, so our probability contours would be incorrect. Giant planet multiplicity around M-dwarfs is not well understood, but since giant planet occurrence is believed to be small \citep{Bonfils13} the multiplicity rate of giant planets around M-dwarfs is likely also small. Presently, there are no known systems with two planets larger than Jupiter orbiting one M-dwarf. Even in cases with two large planets, one planet will dominate the RV signal. For example, OGLE-2006-BLG-109L contains a $0.73 \pm 0.06 M_J$ planet at $2.3 \pm 0.5$ AU and a $0.27 \pm 0.02 M_J$ planet (slightly less massive than Saturn) at $4.5^{+2.1}_{-1.0}$ AU \citep{Gaudi08, Bennett10}. In this case the Doppler amplitude of the inner planet would be a factor of 3.3 larger than the Doppler amplitude of the outer planet. Similarly, an external observer of the solar system would observe an RV signal from Jupiter 4.5 times larger than that of Saturn. Thus we neglect this possible source of error.

We then claim that the likelihood of the existence of a giant planet given the nondetection of an RV acceleration is 
\begin{equation}
P(\mbox{planet}|\mbox{ND}) = f_{pl}(1-\eta_{pl,\star})
\label{FNP} 
\end{equation}
where $\eta_{pl, \star}$ is the probability of detecting a giant planet around a given star as a function of planet mass and orbital semimajor axis, estimated by simulating observations of injected planets. The true probability of missing a planet depends on the true giant planet occurrence rate and the planet distribution function. We can determine this value directly if the underlying planet distribution function (Equation \ref{DPL}) is assumed. By counting the observed trends and analyzing our RV detection efficiencies for each star as a function of mass and separation, we can determine the number of missed planets. We find our final result is not a strong function of mass index $\alpha$ or semimajor axis index $\beta$ (see \textsection\ref{PL}).

\subsection{False Positives}
\label{FP}
\subsubsection{Multiple Planets}
In some cases, observed accelerations may not be induced by the orbit of a giant planet. If two smaller planets are orbiting one star, when they are both on the same side of the star their RV signals would constructively interfere, giving the appearance of a giant planet where none exists. Again, multiplicity rates of large planets are unknown for these small stars but are likely small; we again neglect this effect as a possible source of error. This is a reasonable assumption even if the multiplicity rate of gas giant planets around M-dwarfs was much larger than currently expected. Both the orientation of the system and the relative positioning of the planets during our observations is random. Therefore, it is equally likely that multiple planets would be in the ``constructive'' or ``destructive'' phase of their orbits. Thus, similar numbers of false additional planets would be added to our sample as missed true planets. 

\subsubsection{Secular Acceleration}
A false positive can also be caused by secular acceleration. When a high proper motion star moves quickly relative to the Sun, its peculiar velocity vector changes direction in time, causing the star's systemic radial velocity to increase. For a star with proper motion $\mu$ at a distance $d$ the magnitude of this effect is, to first order,
\begin{equation}
\dot\gamma = 23.0 \textrm{ cm s}^{-1} \textrm{ yr}^{-1} \bigg(\frac{d}{10\textrm{ pc}}\bigg)\bigg(\frac{\mu}{1 \textrm{ arcsec yr}^{-1}}\bigg)^2.
\end{equation}
The secular acceleration $\dot\gamma$ is always positive, so that the star's radial velocity only increases because of this effect. For several nearby stars secular acceleration is large enough to create an apparent acceleration or cause an astrophysical RV acceleration to be incorrectly measured. For example, Barnard's star has a secular acceleration of $4.515 \pm 0.002$ m s$^{-1}$ yr$^{-1}$ \citep{Choi13}, larger than all of our observed accelerations. Fortunately, the magnitude of the secular acceleration can be precisely quantified if the star's distance and proper motion are known. All of our stars have measured proper motions and parallaxes, so we can determine the expected secular contribution. This acceleration is subtracted from the observed radial velocity automatically by the CPS RV pipeline \citep{Howard10}, so this potential source of error is automatically accounted for in our data. Moreover, none of our observed accelerations are consistent with what would be expected from secular acceleration alone. 

\subsubsection{Magnetic Activity}
\label{MA}
Magnetic activity on a star can cause a false positive: rotating active regions can affect the shape of the observed spectral lines and thus the apparent RV \citep{Gray88}. A magnetic cycle can occur over years and hide or mimic a radial velocity signal. We denote the fraction of stars with a magnetically-induced acceleration as $f_A$. \cite{GomesdaSilva12} claim six stars from their sample of 27 M-dwarfs with variability ($22\% \pm 9\%$) have RVs induced by magnetic activity. We are interested in the converse (how many trends are induced by variability?) but their result suggests $f_A$ may be significant. To determine $f_A$, we review all 165 M-dwarfs observed by the CPS team, both as part of this survey and as part of the M2K survey \citep{Apps10, Fischer12}. Between these two programs, there are a total of 34 systems with RV trends. We analyze the $S_{HK}$ values for these stars and find the RV correlates with $S_{HK}$ with a correlation coefficient $|r| > 0.5$ in 7 cases, suggesting $20.6\% \pm 7.8\%$ of long-term RV trends may be magnetically induced. We adopt this value as $f_A$. Even if the true value for $f_A$ is a factor of two larger, it would decrease our planet occurrence rate from $f_{pl} = 6.5\%$ to only $f_{pl} = 4.9\%$, still within our uncertainties.

\subsubsection{Brown Dwarfs}
\label{BD}
Our adaptive optics search is sensitive to all stellar-mass companions, but only to the most massive brown dwarfs. We can detect brown dwarfs larger than approximately $50 M_J$, although this number varies from target to target. For each target, we can determine the fraction of brown dwarfs we would expect to detect by our adaptive optics imaging, given the assumption that a trend was caused by a brown dwarf. We call this efficiency $\eta_{BD}$. Here, we assume a form for the brown dwarf mass function where $dn/d\log(m) \propto m^{0.4 \pm 0.2}$ \citep{Pena-Ramirez12}. Thus we can estimate the likelihood of detecting a brown dwarf around a star in our sample, given that a brown dwarf exists. To estimate the probability a brown dwarf exists, we use the result of \cite{Dieterich12}, who, through an HST/NICMOS snapshot program, estimate that $f_{BD} = 2.3^{+5.0}_{-0.7}\%$ (at $1\sigma$) of M-dwarfs have an L or T companion between 10 and 70 AU. This is consistent with the result of \cite{Metchev09}, who estimate a brown dwarf companion frequency of $f_{BD} = 3.2^{+3.1}_{-2.7}\%$ (at $2\sigma$) around solar-type (FGK) stars.

\subsubsection{White Dwarfs}
\label{WD}
Compact stellar remnants are often faint and such binary companions can evade direct detection, especially when the compact object is cool ($T < 4000 K$) so that the infrared light is dominated by the primary star \citep{Crepp13b}. Since our targets are all M-dwarfs, it is not unreasonable to expect that some may have formed as lower mass companions in binary systems with the higher mass object having evolved off the main sequence to become a white dwarf.  \cite{Napiwotzki09} combine observations of local white dwarfs with galactic structure models and find that in the thin disk there is a white dwarf number density of $n_{WD} = 2.9 \times 10^{-3}$ pc$^{-3}$. From an analysis of PanSTAARS data, \cite{Wheeler12} estimate 20\% of all white dwarfs have an M-dwarf companion ($f_{M|WD}$), somewhat larger than the 12\% found by \cite{Napiwotzki09}. Considering the measurement by \cite{Chang11} of $n_\star = 0.030 \pm 0.002$ stars per cubic parsec, and that approximately 70\% of all stars are M-dwarfs \citep[$f_{M|\star}$,][]{Henry06}, we can determine the fraction of M-dwarfs in the thin disk with a white dwarf companion, a number we define as $f_{WD}$. If we take $f_{M|WD} = 0.16 \pm 0.04$, we find that 
\begin{equation}
f_{WD} = \frac{n_{WD}f_{M|WD}}{n_\star f_{M|\star}} = 2.2\% \pm 0.5\%,
\end{equation}
where the error is dominated by the uncertainty in $f_{M|WD}$.

By combining the false positive events from \textsection\ref{MA}, \textsection\ref{BD}, and \textsection\ref{WD}, we conclude that given the existence of a trend in our data, the likelihood it is caused by a giant planet is 
\begin{equation}
P(\mbox{planet}|\mbox{trend}) = (1-f_A)[1-f_{BD}(1-\eta_{BD, \star})](1-f_{WD})
\label{FPP}
\end{equation}

\subsection{Determining $f_{pl}$}
\label{Methods}

We determine the giant planet occurrence rate, $f_{pl}$ by combining our estimate of the number of false positives and false negatives with the number of observed accelerations. Specifically, the occurrence of giant planets is given by Eq. \ref{ResultEq} if the number of observed accelerations is known, along with the probability of a false negative or false positive in our sample. These probabilities are defined by Equations \ref{FNP} and \ref{FPP}, respectively. 

For each star in our sample, we use our map of giant companion detectability (e.g. Fig. \ref{WrightRV}) to estimate our efficiencies, $\eta_{BD}$ and $\eta_{pl}$. We measure the total planet fraction, $f_{pl}$ and its uncertainty through a Monte Carlo experiment. For each trial, we establish an expected number of observed accelerations, drawing from a binomial distribution with $n=111$ and $p = 4/111$, representing the most likely underlying distribution behind our observed sample. In practice, we draw from our star list 111 times, with replacement, to determine a stellar sample. We then draw randomly from our previously defined distributions to estimate $f_A$, $f_{BD}$ and $f_{WD}$. These values are sufficient to calculate the probability an observed acceleration is caused by a false positive astrophysical event. In cases where known planets with masses $m > 1 M_J$ exist in our sample, we include their presence in our calculation of $f_{pl}$. 

The derivative of the RV acceleration (the ``jerk'') for Gl\,849 is nonzero, so we can use the RV information to fit a two-planet model to this system, instead of a planet plus a linear acceleration (see Appendix). We find the inner planet to have a mass $m\sin i = 0.90 \pm 0.05 M_J$ with a period of $5.24 \pm 0.07$ years, and the outer planet to have a mass $m \sin i = 0.70 \pm 0.31 M_J$ with a period of $19.3^{+17.1}_{-5.9}$ years. More data is needed to determine the exact parameters of the orbit of Gl\,849c, but from the existing RV information we can determine the probability each planet has a mass $m > 1 M_J$. The exact value depends on the planet mass distribution function; assuming each orientation has equal probability (so that $\alpha=-1$) we find probabilities of 0.577 and 0.419, respectively. Following the method of \cite{Ho11}, we find changing the distribution function changes these values by less than 10\%. 


Since we know the region of mass-separation parameter space to which we are sensitive to planets for each star, we can self consistently estimate the planet frequency in this parameter space. We then assume the result from \cite{Cumming08}, who find the power-law indices (Eq. \ref{DPL}) of $\alpha = -0.31 \pm 0.20$ and $\beta = 0.39 \pm 0.15$. We randomly select values for $\alpha$ and $\beta$ from these distributions and use our detection efficiencies to determine the number of false negative missed planets in our sample. Through Equation \ref{ResultEq}, we then have enough information to estimate the planet fraction as a function of each parameter. By repeating this process many times, varying each of our assumed parameters, we can measure the overall planet fraction and its uncertainty.

\section{Results and Discussion}
\label{Results}
\subsection{The Frequency of Giant Planets}
\label{R1}

\begin{figure}[htbp]
\centerline{\includegraphics[width=0.5\textwidth]{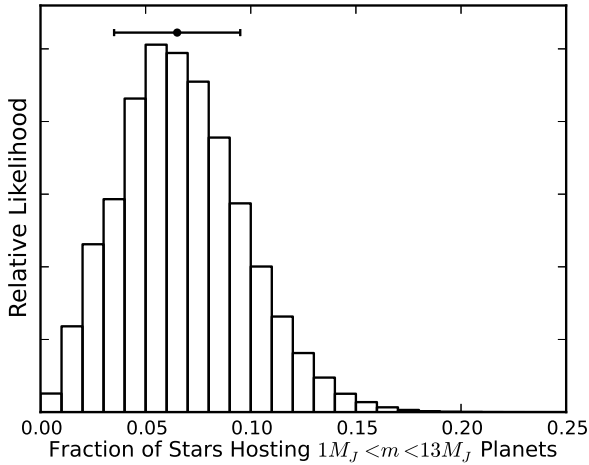}}
\caption{ Giant planet occurrence for our sample of 111 nearby M-dwarfs. We find $6.5\% \pm 3.0\%$ of M-dwarfs host a planet with mass $1 M_J < m < 13 M_J$ and $0 < a < 20$ AU.}
\label{HistoResult}
\end{figure}

Given an observed trend, we can estimate the likelihood the signal is caused by a massive planet. By analyzing our 111 targets as described in \textsection\ref{Methods}, we recover a distribution in giant planet occurrence as shown in Fig. \ref{HistoResult}. We find from this analysis that $6.5\% \pm 3.0\%$ of all M-dwarfs host a giant planet with a semimajor axis smaller than 20 AU. This number is lower than previous studies of higher-mass stars. \cite{Bowler10} find $24^{+8}_{-7}\%$ of ``retired'' A stars host Jupiter-mass planets within 3 AU, while \cite{Cumming08} find that $f_{pl} = 10\% \pm 1\%$ of FGK stars host Jupiter-mass planets within 20 AU. 

If we consider multiplicity in situations where we have a giant planet and an RV acceleration (or in the case of Gl\,849, two giant planets), then we measure a giant planet occurrence rate of $0.083 \pm 0.019$ giant planets per star. To estimate this, we repeat the calcuations of the previous section, but count known giant planets separate from observed accelerations in the cases when we observe both a planet and a ``trend.'' This number does implicitly assume that observed accelerations are caused by the motion of one giant planet, not a combination of multiple planets in motion. The multiplicity rate of giant planets around M-dwarfs appears to be lower than the multiplicity rate of small planets, such as those detected by \textit{Kepler} \citep{Youdin11}.

Our result is consistent with the result of microlensing surveys of M-dwarfs, which suggest a total occurrence rate of $0.09^{+0.03}_{-0.05}$ giant planets per star in the range $1 M_J < m < 10 M_J$ and $0.5$ AU $< a < $20 AU \cite{Cassan12}. However, the power-law distribution determined by the microlensing studies is considerably different than the \cite{Cumming08} distribution assumed here. We discuss this further and constrain $\alpha$ and $\beta$ in \textsection \ref{muL}.

This is the first study using observed RV accelerations to estimate the giant planet occurrence rate. However, previous RV studies have discussed the presence or nondetection of RV accelerations in their analysis. For example, \cite{Endl03} mentioned all RV accelerations in their sample are likely the cause of stellar binaries. Our observations are generally more precise than theirs, as we detect some planets that they miss (such as Gl 436 and Gl 849).

\cite{Bonfils13} detect 15 long-term accelerations in their sample of 102 southern M-dwarfs. Some of these can be attributed to long period binary companions (such as Gl\,250B and Gl\,618A). Of the stars where we detect an RV acceleration, only one (Gl\,849) is in the HARPS sample; these authors also detect an acceleration. \cite{Bonfils13} also detect an acceleration around Gl\,699 (Barnard's star) that we do not detect. Such an acceleration has also not been found by other studies: \cite{Choi13} claim the RV of Barnard's star is increasing at $4.515 \pm 0.002$ m s$^{-1}$ yr$^{-1}$, consistent with the expected secular acceleration but inconsistent with the $-3.043 \pm 0.646$ m s$^{-1}$ yr$^{-1}$ acceleration observed by \cite{Bonfils13}. With more observations over a longer time baseline, this discrepancy will be resolved.

\subsection{Potential Missed Binary Stars}
\label{PMBS}
We only collect AO images for systems with long term RV accelerations. For these accelerations to be observable, the orbiting companion must have a component of its movement along our line of sight so that the radial velocity changes during an orbit. A giant planet would be missed if it was in a near face-on orbit, such that the star's reflex motion was primarily in the plane of the sky. Such systems are accounted for in our detectability calculations (Fig. \ref{Detectability}), as we have determined the probability of detecting a planet's RV acceleration as a function of its mass and separation, marginalized over all other orbital parameters. These calculations do not, however, account for the possible presence of close stellar binary companions in face-on orbits. Although less common than edge-on systems, any missed binary systems that we have not rejected from our sample would cause our planet occurrence rate to be artificially low (assuming these systems could not form dynamically stable planets). Close binaries would be observable as double-lined spectroscopic systems (SB2s) in the CPS data, while wider binary pairs would be easily imaged by AO systems.

The RV sample was originally selected to reject systems with known binary companions within 2 arcseconds. We would expect companions with a flux ratio larger than 0.01 ($\Delta V = 5$) to be detected as binaries \citep{Robinson07}. For our brightest targets, this would correspond to M6 dwarfs and brighter. As the cutoff for hydrogen burning is the M6 spectral class \citep{Luhman12}, we would expect all close stellar-mass binaries to be removed from our HIRES observations. 

To determine how many missed binaries are in our sample, we simulate a population of binary companions to M-dwarfs. We create binary companions such that their semimajor axes are assigned following the observed distribution found by \cite{Fischer92}. We randomly assign the other orbital parameters and determine there is a $41.8\% \pm 0.3\%$ chance a binary companion in our sample around a random star would have a projected separation smaller than two arcseconds. Thus, considering \cite{Fischer92} find $42\% \pm 9\%$ of local M-dwarfs are in binary or multiple systems, we would expect to have a total of $24 \pm 6$ binary systems in our sample, which originally contained 137 stars before the removal of known binaries. As we actually observe 22 binary systems (containing 26 stars), this result is consistent with our expectation.

We then determine the radial velocity each simulated binary star would induce on our host companion. For each binary that induces a measurable acceleration on the host star, we simulate imaging observations to determine the probability this binary companion would be detected in either our AO survey or, for very wide separation binaries, a seeing-limited ground based survey such as 2MASS. By applying our joint AO/seeing-limited contrast curves, we find that if a binary star system in our survey induces an RV acceleration, we would have a $96.0\% \pm 0.4\%$ chance of imaging this binary companion. Therefore the probability that one or more of our observed accelerations is caused by a ``missed'' binary companion is negligible and this possibility does not significantly affect our results.

\subsection{Dependence on Stellar Mass}
\label{MassDep}
Previous RV studies have found a correlation between stellar mass and giant planet occurrence at $a < 2.5$ AU, with more massive stars more likely to host giant planets \citep{Johnson10a}. To test this relation inside the M-dwarf spectral class, we analyzed the high mass stars separately from the low mass stars in our sample. From our best fit masses, half of our sample is more massive than $M = 0.41 M_\odot$. We thus use this value as a dividing line to separate our sample into two groups. Our masses have typical uncertainties of $10\%$, so for each star, given its mass and uncertainty, we determine the probability it is larger or smaller than $0.41 M_\odot$ assuming normally distributed errors. We then use that value as a weighting factor to assign a probability for each individual star to reside in our high mass or low mass bin, and then repeat our analysis for each individual subsample.

\begin{figure}[htbp]
\centerline{\includegraphics[width=0.5\textwidth]{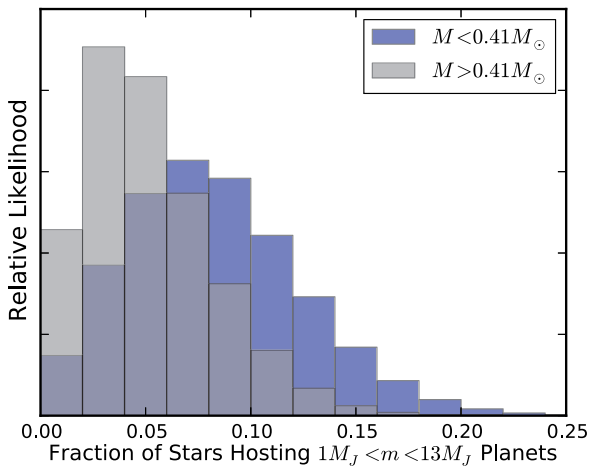}}
\caption{ Planet occurrence for a low mass subsample (blue) and a high mass subsample (gray) of M dwarfs. Both subsets have nearly similar giant planet occurrence rates, suggesting planet occurrence may not depend strongly on stellar mass within the M spectral class. A larger sample is required to determine if the lack of difference in occurrence rates is astrophysical or statistical variance.
  }
\label{Mass}
\end{figure}

We find an occurrence rate for the high mass subsample of $4.8 \pm 3.3\%$ and for the low mass subsample of $7.9 \pm 4.2\%$ (Fig. \ref{Mass}). \cite{Johnson10a} find planet occurrence is correlated with stellar mass such that $f_{pl} \propto 10^{(1.2\pm 0.2)\textrm{[Fe/H]}}M_\star^{(1.0\pm 0.3)}$. The average star in our high-mass sample has a mass of 0.5 $M_\odot$ while the average star in our low-mass sample has a mass of 0.3 $M_\odot$, so we would expect the high-mass subsample to have an occurrence rate larger than the low-mass sample by a factor of 1.67. We find the true occurrence rate to change by a factor of $0.61 \pm 0.87$ in moving from the lower-mass to higher-mass bin. This is inconsistent with the expected result from \cite{Johnson10a}, but the difference between the two bins is not significantly different from zero. A larger sample is required to determine if the small difference between these two populations of M-dwarfs is real or the result of a statistical anomaly. However, our result is lower than the \cite{Cumming08} result for FGK stars, that $f_{pl} = 10\% \pm 1\%$ of FGK stars host Jupiter-mass planets within 20 AU. This difference is consistent with the \cite{Johnson10a} correlation between stellar mass and planet occurrence.

\subsection{Dependence on Metallicity}
\label{MetalDep}
Previous RV studies of giant planets have also found evidence for a correlation between planet occurrence and metallicity \citep{Fischer05, Johnson10a}. To test if this correlation holds for more distant planets, we again split our sample into two, using the same method from the previous section. In this case, we use [Fe/H] $= -0.10$, the sample median metallicity, as the dividing line for our subsamples. We assume all stars have metallicity uncertainties of 0.17 dex, consistent with the scatter expected from the \cite{Neves12} empirical relation. Again, we assume Gaussian errors to determine the probability each star is in a specific subsample. We then repeat our analysis on both groups.

\begin{figure}[htbp]
\centerline{\includegraphics[width=0.5\textwidth]{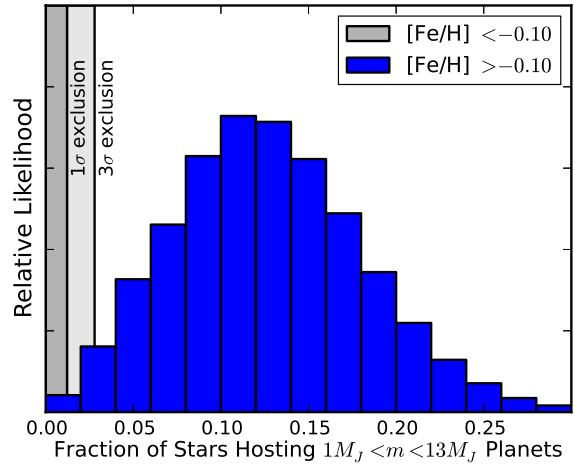}}
\caption{ Planet occurrence for a high metallicity subsample (blue) and 1$\sigma$ and 3$\sigma$ exclusion regions for a low metallicity subsample (gray) of M dwarfs. In the low metallicity subsample, we are able to rule out planet occurrence rates larger than 1.2\% at $1\sigma$ and 2.8\% at $3\sigma$, represented by the labeled vertical lines. The high metallicity sample has a significantly higher occurrence rate than the low metallicity sample, similar to the phenomenon observed for RV-detected planets at smaller separations. 
  }
\label{Metals}
\end{figure}
In the high-metallicity subsample, we find an occurrence rate such that $12.4 \pm 5.4\%$ of M-dwarfs host giant planets. In the low metallicity sample the occurrence rate drops to $0.96 \pm 0.51\%$. In Fig. \ref{Metals} we plot a histogram of our posterior distribution of planet occurrence for our high-metallicity subsample. Vertical lines represent (from left to right) 1$\sigma$ and 3$\sigma$ upper limits on the planet occurrence rate for the low-metallicity subsample. From these distributions, the giant planet occurrence rate for metal-rich stars has only a $2.4\%$ probability of being lower than the $3\sigma$ upper limit on the planet occurrence rate for metal-poor stars. The difference between these subsamples may be suggestive of the same effect seen for RV-confirmed planets within 2.5 AU \citep{JohnsonApps09, KOI254}.

An increase in the planet occurrence rate with metallicity for planets beyond a few AU may suggest giant planets in wide orbits are commonly formed by the same processes as the RV giant planet population. This study will be facilitated by the development of reliable spectroscopic metallicity measurements \citep{RA10}.

\subsection{The Stellar Mass-Metallicity Plane}
We can quantify our giant planet occurrence rate with respect to stellar mass and metallicity. Such an approach has been undertaken for planets with $a < 2.5$ AU orbiting stars of all spectral types previously \citep{Johnson10a}; we follow the techniques of these authors but confine ourselves to strictly giant planets in the range $0 < a < 20$ AU orbiting stars of the M-dwarf spectral class.

We assume that stellar mass and metallicity produce separate effects on the giant planet occurrence rate, so that the fraction of stars with planets as a function of mass and metallicity can be written as a double power-law,
\begin{equation}
f(M,F) = CM^a10^{bF},
\end{equation}
where $C$, $a$, and $b$ are constants, $M \equiv M/M\odot$, and $F \equiv$ [Fe/H].

In this analysis, we have a binary result: a star either has a giant planet, detectable as an RV acceleration or closed orbit, or it does not. Therefore, each of the $N$ stars in our sample represents a Bernoulli trial. Given $T$ total observed giant planets, if we assume the probability of a Doppler detection of a giant planet around any given star $i$ is $f(M_i, F_i)$, then by Bayes' theorem, the probability of a given model $X$ given our data $d$ is
\begin{equation}
P(X|d) \propto P(X) \prod_i^T f(M_i,F_i) \times \prod_j^{N-T} [1- f(M_j,F_j)].
\label{Likeli}
\end{equation}

Our measurements of stellar masses and metallicities are imperfect. Therefore, we treat the masses and metallicities of these stars as probability distributions. We consider each star's mass and metallicity distribution to be a two-dimensional Gaussian with mean {$M_i,F_i$} and standard deviation {$\sigma_{M,i},\sigma_{F,i}$ and call this term $p$. In this case, the predicted planet fraction for a star with mass $M_i$ and metallicity $F_i$ is
\begin{equation}
f(M_i,F_i) = \int \int p(M_i, F_i)f(M,F)dM dF.
\end{equation}
We can thus apply Eq. \ref{Likeli} with varying parameters, $X = {C, a, b}$, to maximize $\mathcal{L}$ conditioned on the data. We elect to use uniform priors, instead of applying the results of previous studies as a prior. \cite{Johnson10a} and \cite{Mortier13} study a sample of stars including all stellar types F to M, so their results may not represent our population well. More recent studies, such as \cite{Neves13}, are restricted to M-dwarfs. However, while their techniques are similar, they only attempt to constrain metallicity, implicitly assuming $a=0$. Additionally on of the three detected planets in their sample is a planet smaller than Jupiter around a metal-poor star. As our sample is limited to planets larger than Jupiter, the resultant distribution found by these authors may not be representative of the population of giant planets ($m > 1 M_J$).}

\begin{figure}[htbp]
\centerline{\includegraphics[width=0.5\textwidth]{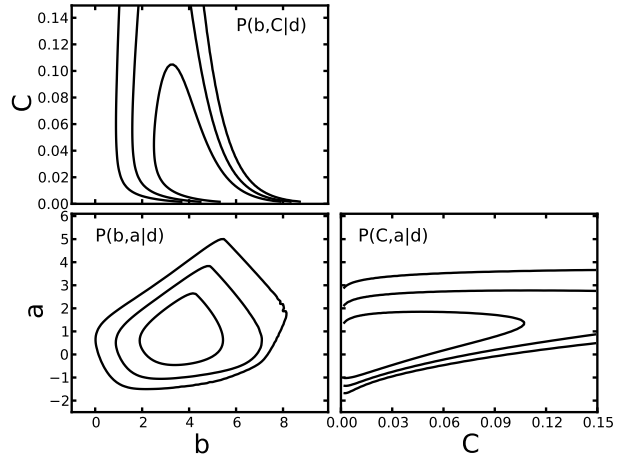}}
\caption{Marginal posterior pdfs for the planet population model conditioned on our M-dwarf data. We find, as by other methods in previous sections, that giant planet occurrence is a strong function of stellar metallicity, but may not depend strongly on stellar mass inside of the M spectral class. 
  }
\label{Covar}
\end{figure}

We find our giant planet fraction is described by the distribution function
\begin{equation}
f(M,F) = 0.039^{+0.056}_{-0.028} M^{0.8^{+1.1}_{-0.9}} 10^{(3.8 \pm 1.2)F}.
\end{equation}
The $1\sigma$ confidence interval for $C$ is highly skewed, while the other two parameters are approximately normally distributed. In Fig. \ref{Covar}, we plot the marginal posterior probability distribution functions for each pair of parameters. Perhaps not surprisingly, we find a covariance between $C$ and $b$. Because our metallicity parameter $b$ is so steep, small changes in $b$ must cause changes in $C$ to keep the giant planet fraction consistent at a given metallicity.

Our results are steeper in $b$ than \cite{Neves13}, although the giant planet occurrence rates at [Fe/H] $\sim 0.1$ are consistent between the two studies. This is likely due to the inclusion of a planet with a minimum mass of 0.7 Jupiter masses in the ``Jovian'' sample of these authors. This planet orbits a star with a metallicity [Fe/H] $= -0.19 \pm 0.08$, flattening the distribution with metallicity. The fact remains that, while the metallicity distribution of field stars is centered near [Fe/H] $=0.0$ with a standard deviation of 0.13 dex, there are presently no giant planets orbiting M-dwarfs with measured metallicities smaller than +0.08 in either the HARPS or HIRES sample. The giant planet distribution function must therefore be a strong function of stellar metallicity. Moreover, it is essential to develop improved methods to measure metallicities of low-mass stars, such as the techniques developed by \cite{RojasAyala12} and \cite{Mann13}.

\subsection{The Effect of Distant Binary Companions}

In the above analysis, we neglect binary stars where a test particle at 30 AU would be in an unstable orbit, but include 14 binaries at wider separations. Although these systems formally allow stable orbits, \cite{Kaib13} suggest these orbits can change significantly over time. Because the binary pair is weakly bound, interactions with the galactic tidal field or nearby passing stars can vary the binary orbit. The binary can then strongly perturb formerly stable planetary companions, potentially resulting in the ejection of planets from the system within 5 Gyr, our estimated age for the M-dwarfs in our sample. None of our 10 wide binary systems show evidence for an RV acceleration, providing weak but tantalizing evidence in favor of this theory. If we repeat our analysis but neglect these stars as potential hosting systems, we find that $7.4\% \pm 3.3\%$ of single stars host giant planets, compared to $6.5\% \pm 3.0\%$ of our full sample. With zero detections in a sample of 14 wide binaries, we can only place an upper limit of $f_{pl} \leq 0.20$ at $95\%$ confidence on the occurrence rate of giant planets in wide binary systems. With more observations of stars with wide binary companions, the occurrence rate of planets orbiting true field stars can be compared to the rate for wide binaries.

\subsection{Sensitivity to Power-Law Parameters}
\label{PL}
\begin{figure}[htbp]
\centerline{\includegraphics[width=0.5\textwidth]{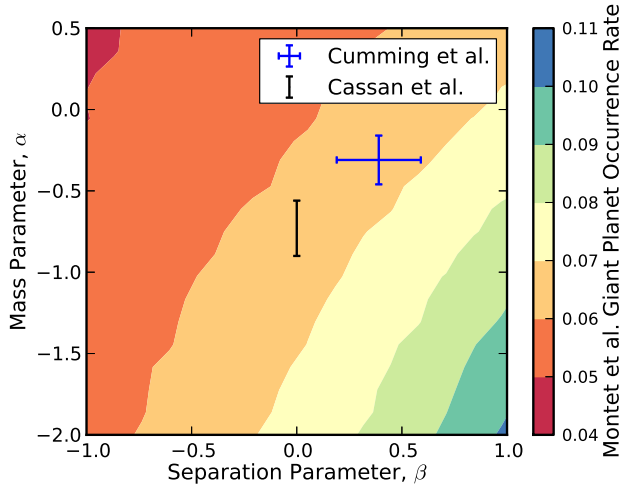}}
\caption{Calculated giant planet occurrence rate, $f_{pl}$, as a function of the mass parameter index $\alpha$ and separation parameter index $\beta$. There is not a strong dependence on $\alpha$ or $\beta$; selecting $\alpha < -1.0$ and $\beta > 0.5$ is required to affect our result at more than the $1\sigma$ level. Labeled points include the \cite{Cumming08} result for FGK stars, with $\alpha = -0.31 \pm 0.15$ and $\beta = 0.39 \pm 0.15$, and the microlensing result of \cite{Cassan12}, who find $\alpha = -0.73 \pm 0.17$ and assume $\beta \equiv 0$.
  }
\label{AlpBetOcc}
\end{figure}

The result for $f_{pl}$ is dependent on the exact parameters of the planetary distribution function, as that function determines the number of missed (false negative) planets in our sample. To quantify the dependence of the planetary occurrence rate on our choice of $\alpha$ and $\beta$ we repeat our analysis over a grid of values for $\alpha$ and $\beta$. The giant planet occurrence rate as a function of these two parameters is shown in Fig. \ref{AlpBetOcc}. We find that there is only a weak relation between $\alpha$ and $f_{pl}$ in the range $-2.0 < \alpha < 0.5$, where we might reasonably expect $\alpha$ to reside. $f_{pl}$ depends more strongly on $\beta$, but our overall result does not change by more than $1\sigma$ by selecting any $\beta$ in the range $-1.0 < \beta < 1.0$ for a given $\alpha$. Selecting any $\alpha$ or $\beta$ over this range affects our final result by less than a factor of two.

From our sample of targets alone, we are unable to place constraints on acceptable values of $\alpha$ and $\beta$. To constrain $\alpha$ and $\beta$, the occurrence rate of giant planets at a given mass or separation is required. We have determined the bulk occurrence rate of planets, but cannot uniquely determine their properties. With continued observations, as our RV accelerations ``turn over'' and become closed orbits, we will be able to determine the exact locations of giant planets around M-dwarfs and constrain the power-law parameters. Alternatively, we can constrain $\alpha$ and $\beta$ by combining our results with those from microlensing observations.

\subsection{Comparison with Microlensing Results}
\label{muL}
 
In \textsection\ref{PL}, we showed that our bulk occurrence rate is not a strong function of $\alpha$ and $\beta$. However, the types and locations of our planets is a function of these parameters: if $\alpha$ is large, then most of our observed trends must be caused by large planets in wide orbits. Since microlensing results are most sensitive at projected separations corresponding to the Einstein radius, where $R_E \sim 3.5 \textrm{AU} (M_\star/M_\odot)^{1/2}$, we can compare our results to microlensing planet occurrence studies. As our results will only be consistent with microlensing estimates of the planet occurrence rate at the Einstein radius for specific values of $\alpha$ and $\beta$, comparisons between the two methods will enable us to constrain $\alpha$ and $\beta$. 

To compare the two sets of results, we assume the population of M-dwarfs observed by microlensing studies is similar to that targeted by RV surveys in the local neighborhood.  We find evidence for a correlation between giant planet frequency and metallicity in our sample, similar to that found by previous RV analyses of planets with $a < 2.5$ AU \citep{Fischer05, JohnsonApps09}. M-dwarfs studied by microlensing are at distances larger than 1 kpc and in the direction of the galactic bulge, along the galactic metallicity gradient \citep{Rolleston00}. Measurements of the metallicity of Cepheids suggest the iron content in the disk varies such that $d\textrm{[Fe/H]}/dr = -0.051 \pm 0.004$ dex kpc$^{-1}$ between 5 and 17 kpc from the galactic center \citep{Pedicelli09}. Thus, the microlensing M-dwarfs may be more metal-rich than stars in the local neighborhood, so $f_{pl}$ may be larger for the microlensing population than the RV population. Without spectra of galactic stellar planet-hosting lenses their true stellar properties are unknown. Programs dedicated to collecting spectra of galactic stellar planet-hosting lenses would greatly inform our knowledge of these stars and their planets.

If we assume the planet mass distribution function of \cite{Cumming08}, then from our analysis we would expect microlensing studies to measure a planet occurrence rate $f_{pl} = 0.056 \pm 0.023$ bound Jupiter-mass planets per star by analyzing signals from planets near the Einstein radius. \cite{Cassan12} claim an occurrence rate of $10^{-0.62\pm0.22}$ ($0.24^{+0.16}_{-0.10}$) Saturn-mass planets at this separation. If we scale this occurrence rate to Jupiter-mass planets following the mass index observed in microlensing studies, $\alpha = -0.73 \pm 0.17$, then the observed microlensing density of Jupiter mass planets would be $0.101 \pm 0.016$ planets per star, different from our expectation at $1.6\sigma$.  If (and only if) the two populations have intrinsically similar occurrence rates of giant planets, then the difference between the number of planets found must be due to a planet distribution different from the one used by \cite{Cumming08}. As the RV planet distribution was developed from an analysis of FGK stars, while the microlensing population generally consists of M dwarfs that may be preferentially metal-rich compared to stars in the local neighborhood, it may not be surprising if the RV planet population is intrinsically different from the microlensing planet population.

\subsubsection{Joint Constraints on $\alpha$}

We depart from our previously assumed values of $\alpha$ and $\beta$ to determine what values of $\alpha$ and $\beta$ satisfy both our observed RV accelerations and the results of \cite{Cassan12}. We assume the planet occurrence rate presented by \cite{Cassan12} is representative of the planet population at the Einstein radius. Moreover, we assume planet orbital semimajor axes are distributed uniformly in logarithmic space following \"Opik's Law ($\beta = 0$), as microlensing studies assume. This is slightly shallower than what is observed in the RV planet population ($\beta = 0.39 \pm 0.15$), but since the RV population of giant planets likely underwent considerable migration this may be a reasonable assumption. We then vary $\alpha$, and for each value determine the space density of planets at 2.5 AU. We then compare our expected result to the result from \cite{Cassan12}, which we scale to Jupiter-mass planets according to our $\alpha$ parameter. We finally require $\alpha < 0$: despite the uncertainties in this mass parameter, previous studies agree that around M dwarfs, small planets are more common than massive planets \citep{Swift13, Morton13}. 

\begin{figure}[htbp]
\centerline{\includegraphics[width=0.5\textwidth]{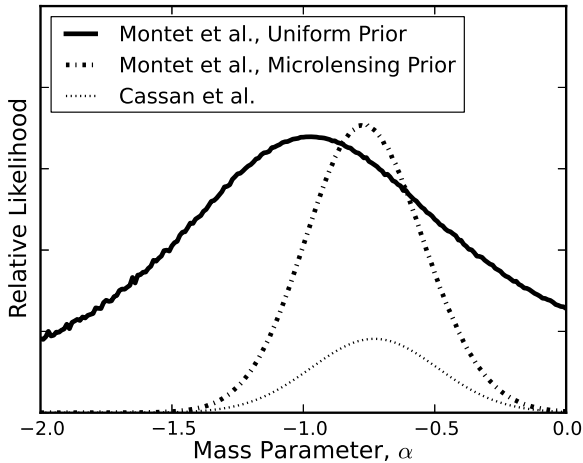}}
\caption{Relative likelihood values for the mass parameter $\alpha$, assuming the planets in our sample and microlensing systems are members of the same population. We find a maximum likelihood value of $\alpha = -0.94 \pm 0.56$, consistent with values of $\alpha$ found from analyses of microlensing planets but steeper than previous RV results for FGK stars at $1.1 \sigma$. This result may suggest the planet distribution function is different for M stars as compared to higher mass stars. When we include the \cite{Cassan12} result as a prior on our measurement, we find $\alpha=-0.77 \pm 0.22$.
  }
\label{alphaopik}
\end{figure}

We find microlensing results agree with our result for $f_{pl}$ when $\alpha = -0.94 \pm 0.56$ (Fig. \ref{alphaopik}). This result is consistent with the best-fitting values for $\alpha$ found by \cite{Gould10} and \cite{Cassan12}. If we include the \cite{Cassan12} result as a prior in our analysis, we find $\alpha = -0.77 \pm 0.22$. However, while our result agree with microlensing studies, our result for $\alpha$ is different from the \cite{Cumming08} result for FGK stars at $1.1 \sigma$ and significantly different from the \cite{Bowler10} constraints for A stars, which rule out all $\alpha < 0.25$ with 90\% confidence and all $\alpha < 1.75$ with 50\% confidence. Since microlensing predicts a larger number of planets found at the Einstein radius relative to that expected by RV extrapolations, it is not surprising that we find a smaller value for $\alpha$ is required for our result to be consistent with the microlensing results: if the two populations are the same, there must be many low-mass giant planets below the simultaneous RV and imaging detectability limits than high-mass planets above the limits.

\subsubsection{Simultaneous Constraints on $\alpha$ and $\beta$}

\begin{figure}[htbp]
\centerline{\includegraphics[width=0.5\textwidth]{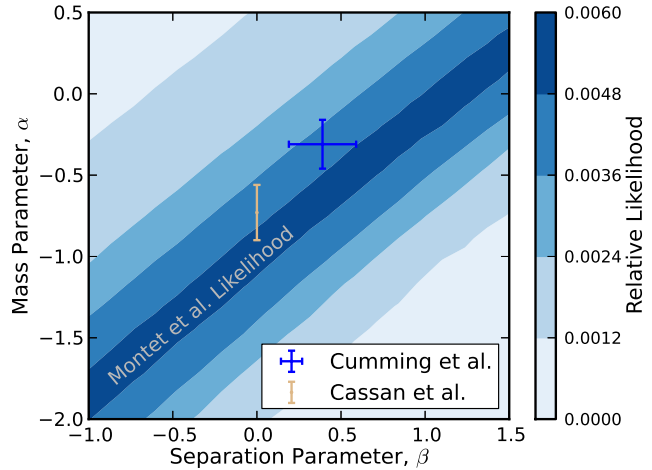}}
\caption{Relative likelihood values for the mass parameter $\alpha$, and separation parameter $\beta$. There is a maximum likelihood contour approximately along the line $\alpha - \beta = -1$, suggesting a relationship between the two parameters required to fit both our result and the microlensing results, assuming the local planets in our sample and microlensing systems are members of the same population. Points included in the plot are the \cite{Cumming08} RV result (blue) and the \cite{Cassan12} microlensing result (cyan), the latter of which assumes an \"Opik's Law value of $\beta = 0$. The small discrepancy between our result and the \cite{Cumming08} result may suggest the planet distribution function may differ between M-dwarfs and FGK stars.
}
\label{alphabeta}
\end{figure}

We are not restricted to \"Opik's Law. We can allow both $\alpha$ and $\beta$ to vary, and compare the normalization of \cite{Cassan12} for Saturn-mass objects at 2.5 AU to our projected planet density at that mass and separation (Fig. \ref{alphabeta}). Performing this exercise, we find the most acceptable values of $\alpha$ and $\beta$ are correlated approximately along the line $\alpha - \beta = -1$. That is, for every 1 dex increase in $\alpha$, $\beta$ must decrease by 1 dex to maintain a reasonable fit to both our result and the microlensing results. 

\subsubsection{A Model-Independent $f_{pl}$}
We can apply these relative likelihood values as priors to the occurrence rate as a function of $\alpha$ and $\beta$ shown in Fig. \ref{AlpBetOcc} to determine an occurrence rate independent of our choices of $\alpha$ and $\beta$, but dependent on the RV and microlensing stars both being representative of similar populations. We assume our separation parameter must be in the range $-1.0 < \beta < 1.0$, consistent with the assumptions from previous microlensing studies, and allow our mass parameter to be any value subject to the constraints of Fig. \ref{alphabeta}. By weighting our occurrence rates found in \textsection\ref{PL} in this manner, we find a most likely occurrence rate of $7.2 \pm 3.1\%$, consistent with that found by assuming the power-law distribution of \cite{Cumming08}. As the measured planet frequency depends on the distribution function parameters, an improved value of the planet occurrence rate, either by this method, microlensing, or through astrometry measured by \textit{Gaia} \citep{Casertano08}, will provide immediate constraints on the distribution function of giant planets. Similarly, improved constraints on the distribution parameters will enable an immediate improvement of the determination of the giant planet occurrence rate.

The \cite{Cumming08} power-law parameters $\alpha$ and $\beta$ are less consistent with our results. This may suggest the planet distribution function around FGK stars is systematically different from the planet distribution function around M-dwarfs. As \cite{Bowler10} find an even larger value for $\alpha$ in their study of retired A stars (excluding all $\alpha < 0$), which matches comparison studies between RV surveys and high-contrast imaging searches \citep{Crepp11}, this possibility is certainly plausible. With additional M-dwarfs targeted by a combination of RV observations with longer time baselines and high-contrast imaging to improve the estimate of the occurrence rate, we will be able to directly probe this possibility.

\section{Summary and Conclusion}
\label{SC}
We have analyzed a collection of 111 nearby M-dwarfs observed in RV surveys with a median time baseline of 11.8 years in a search for long-term RV accelerations. We have developed a new technique to determine the incidence of giant planets in which we target systems with such accelerations using adaptive optics imaging to ``peer beyond the horizon'' set by Doppler time baselines. With a relatively short exposure image using the Keck AO system, we can eliminate the possibility of binary stellar companions and massive brown dwarfs. We conclude with high statistical confidence that accelerations without a directly imaged companion are likely caused by a planet in a wide orbit. 

Accounting for false positive and false negative rates, we find that $6.5 \pm 3.0\%$ of M-dwarfs host a giant planet with mass $1 < m / M_J < 13 $ and semimajor axis $0 < a < 20$ AU, assuming such planets are distributed following the power-law parameters estimated by \cite{Cumming08}. The exact integrated planet occurrence rate does not depend strongly on the distribution function parameters chosen. We find evidence for a correlation between giant planet frequency and stellar metallicity, similar to that observed in the RV-detected planet population. Additional follow-up work confirming this result would suggest giant planets in wide orbits may form in the same way as the RV-detected giant planets. Observations of more stars are needed to determine if a correlation exists between planet occurrence at wide separations and stellar mass inside of the M-dwarf spectral class. 

Our overall occurrence rate is consistent with what might be expected based on the results of microlensing planet search surveys. However, if the giant planet distribution is given as a double power law similar to that found by \cite{Cumming08}, such that $d^2N \propto M^\alpha a^\beta d\ln M d\ln a$, with $\alpha = -0.31 \pm 0.20$ and $\beta = 0.39 \pm 0.15$, where $\alpha$ and $\beta$ are planet distribution power-law indices defined in Eq. \ref{DPL}, then microlensing studies overestimate the giant planet occurrence rate. From our bulk occurrence rate, we determine an expected planet detection rate for microlensing studies which depends on our chosen planet distribution function. By assuming an \"Opik's Law distribution (i.e., flat in $\log a$), the microlensing planet occurrence rate is consistent with our result if the planet population is represented by the power-law $dN \propto m^{-0.94 \pm 0.56} d\log m$. This value for $\alpha$ is consistent with previous M-dwarf studies conducted by microlensing planet search teams \citep{Gould10, Cassan12}. We also find other non-\"Opik distributions can be chosen to simultaneously explain our results and the microlensing results; these fall approximately on the line $\alpha - \beta = -1$. Moreover, an improved estimate of the giant planet occurrence rate, as measured by \textit{Gaia}, can be combined with our results to provide enhanced constraints on $\alpha$ and $\beta$.  

Our knowledge of planets around M-dwarfs has significantly improved in the last few years thanks to both targeted RV searches and high contrast imaging campaigns \citep{Apps10, Bowler12a}. As such surveys continue, they will begin to confirm and characterize planets in wider orbits, pushing into the domain currently only studied by microlensing studies. To directly compare these populations, understanding the properties of host stars to planets found by microlensing will be extremely important; when possible, every effort should be made to collect spectroscopic followup data on microlensing events to determine the physical properties of lens host stars to better understand both the planet population around M-dwarfs and how it changes across the galaxy.

The method developed in this paper can be extended to higher-mass stars with little difficulty. For example, a large sample of K-dwarfs has been observed by the CPS collaboration. This sample is larger, has more observations, and exhibits less astrophysical jitter than our M-dwarf sample; all of these factors improve our ability to detect RV accelerations. However, the stars are more luminous and on average more distant, complicating adaptive optics searches. Care must be taken to ensure low-mass stellar companions are accounted for, as adaptive optics imaging may not be sensitive to all M-dwarf companions to K-dwarfs without longer observations or the use of ADI. In the future, we intend to apply this technique to the CPS K-dwarfs to determine the planet occurrence rate around higher mass stars and compare to the M-dwarfs.

\acknowledgements
Most of the data presented herein were obtained at the W.M. Keck Observatory, which is operated as a scientific partnership among the California Institute of Technology, the University of California and the National Aeronautics and Space Administration. The Observatory was made possible by the generous financial support of the W.M. Keck Foundation. We made use of the SIMBAD database operated at CDS, Strasbourge, France, and NASA's Astrophysics Data System Bibliographic Services. We thank Jon Swift for collecting NIRC2 observations of HIP\,57050. We also thank Brendan Bowler for helpful comments on an early version of this manuscript. B.T.M. is supported by the National Science Foundation Graduate Research Fellowship under Grant No. DGE‐1144469. The TRENDS high-contrast imaging program is supported by NASA Origins grant NNX13AB03G. J.A.J. is supported by generous grants from the David and Lucile Packard Foundation and the Alfred P. Sloan Foundation. B.T.M. would also like to thank the Statistical and Applied Mathematical Sciences Institute; conversations at the June 2013 Modern Statistical and Computational Methods for Analysis of Kepler Data workshop at SAMSI improved the final version of this manuscript.

Finally, the authors wish to recognize and acknowledge the very significant cultural role and reverence that the summit of Mauna Kea has always had within the indigenous Hawaiian community.  We are most fortunate to have the opportunity to conduct observations from this mountain.


\clearpage

\LongTables
\begin{deluxetable*}{l|cccccccccccc}
        \tabletypesize{tiny}
        \tablecolumns{12}
        \tablecaption{M-dwarf stars analyzed in this study}
        \tablehead{\colhead{Star} & \colhead{RA} & \colhead{Dec} & \colhead{Mass ($M_\odot$)} & \colhead{[Fe/H]} & \colhead{Spectral Type} & \colhead{$V$} & \colhead{$V$ Source} & \colhead{d (pc)} }
        \startdata
    Hip 428 & 00:05:10.9 & +45:47:11.6 & 0.53 & -0.07 & M1 &  9.93 & \cite{Gliese91} & 11.25  \\ 
    HD 225213 & 00:05:24.4 & -37:21:26.5 & 0.39 & -0.42 & M1.5 &  8.57 & \cite{Koen10} &  4.34  \\ 
   Hip 1734 & 00:21:56.0 & -31:24:21.8 & 0.55 & 0.09 & M1.5 &  11.1 & \cite{Koen10} & 17.98    \\ 
      Gl 26 & 00:38:59.0 & +30:36:58.5 & 0.43 & 0.02 & M2.5 & 11.2 & \cite{Hog00} &  12.6  \\ 
   Hip 3143 & 00:39:58.8 & -44:15:11.6 & 0.55 & -0.09 & M0.5 &  11.4 & \cite{Koen10} & 23.99 \\ 
      Gl 48 & 01:02:32.2 & +71:40:47.3 & 0.48 & 0.06 & M3 & 10.0 & \cite{Hog00} &  8.24   \\ 
      Gl 49 & 01:02:38.9 & +62:20:42.2 & 0.58 & 0.06 & M1.5 & 9.56 & \cite{Hog00} & 9.96    \\ 
   Hip 5643 & 01:12:30.6 & -16.59.56.3 & 0.13 & -0.43 & M4.5 & 12.1 & \cite{Koen10} & 3.69  \\
   Hip 8051 & 01:43:20.2 & +04:19:18.0 & 0.41 & -0.16 & M2 &  10.9 & \cite{Koen10} & 11.41  \\ 
   Gl 83.1 & 02:00:13.0 & +13:03:07.0 & 0.15 & -0.31 & M4.5 & 12.3 & \cite{Landolt92} & 4.50    \\
  G244-047 & 02:01:35.3 & +63:46:12.1 & 0.48 & 0.07 & M3 & 11.0 & \cite{Hog00} & 12.76    \\ 
      Gl 87 & 02:01:35.3 & +63:46:12.1 & 0.45 & -0.32 & M1.5 & 10.0 & \cite{Koen10} & 10.41 \\ 
  Hip 11048 & 02:22:14.6 & +47:52:48.1 & 0.62 & -0.08 & M0.5 & 9.41 & \cite{Gliese91} & 11.94   \\ 
    Gl 105B & 02:36:15.3 & +06:52:19.1 & 0.27 & -0.10 & M4 & 11.6 & \cite{Jenkins09} & 7.73 \\ 
     Gl 109 & 02:44:15.6 & +25:31:24.1 & 0.35 & -0.18 & M3 & 10.6 & \cite{Koen10} &  7.51    \\ 
  Hip 21556 & 04:37:42.9 & -11:02:19.9 & 0.48 & -0.11 & M1.5 &  10.3 & \cite{Koen10} & 11.10    \\ 
  Gl 179 & 04:52:05.7 & +06:28:35.6 & 0.36 & 0.13 & M3.5 &  12.0 & \cite{Koen10} & 12.29  \\
  Hip 22762 & 04:59:50.0 & -17:46:24.3 & 0.42 & -0.20 & M2 & 10.9 & \cite{Koen10} & 12.12  \\ 
  Hip 23512 & 05:03:20.1 & -17:22:24.7 & 0.27 & -0.25 & M3 & 11.7 & \cite{Koen10} & 9.21   \\
    HD 33793 & 05:11:40.6 & -45:01:06.3 & 0.27 & -0.81 & M1 &  8.85 & \cite{Koen10} &  3.91   \\ 
 Hip 24284 & 05:12:42.2 & +19.39.56.4 & 0.45 & -0.16 & M2 & 10.7 & \cite{Koen10} & 12.29  \\ 
    HD 36395 & 05:31:27.4 & -03:40:38.0 &  0.60 & -0.05 & M1.5 &  7.92 & \cite{Koen10} & 5.66  \\ 
  G097-054 & 05:34:52.1 & +13:52:47.2 & 0.37 & 0.05 & M3.5 &  11.9  & \cite{Kharchenko01} &  12.39  \\ 
  HD 233153 & 05:41:30.7 & +53:29:23.3 & 0.60 & 0.05 & M0.5 &  9.75 & \cite{Gliese91} & 12.44  \\
  Hip 26857 & 05:42:09.3 & +12.29:21.6 & 0.22 & -0.24 & M4 & 11.5 & \cite{Landolt92} & 5.83    \\
   G192-13 & 06:01:11.1 & +59:35:50.8 &  0.27 & -0.11 & M3.5 &  11.7 & \cite{VanAltena95} &  7.93   \\ 
  Hip 29052 & 06:07:43.7 & -25:44:41.5 & 0.30 & -0.22 & M4 & 11.9 & \cite{Koen10} & 11.35   \\ 
     Gl 226 & 06:10:19.8 & +82.06:24.3 & 0.41 & -0.14 & M2 & 10.5 & \cite{Gliese91} & 9.37   \\ 
     Gl 229B & 06:10:34.6 & -21:51:52.7 & 0.58 & -0.07 & M1 & 8.13 & \cite{Koen10} &  5.75  \\
    Gl 250B & 06:52:18.1 & -05:11:24.2 & 0.45 & -0.12 & M2 & 10.1 & \cite{Gliese91}  &  8.71  \\
    HD 265866 & 06:54:49.0 & +33:16:05.4 & 0.35 & -0.03 & M3 & 10.11 & \cite{Hog00} &  5.59  \\ 
     Gl 273 & 07:27:24.5 & +05:13:32.8 & 0.29 & -0.07 & M3.5 & 9.87 & \cite{Koen10} &  3.80   \\ 
  Hip 36338 & 07:28:45.4 & -03:17:53.4 & 0.40 & 0.03 & M3 & 11.4 & \cite{Koen10} & 12.29 \\ 
  Hip 36834 & 07:34:27.4 & +62:56:29.4 & 0.40 & -0.50 & M0.5 & 10.4 & \cite{Hog00} & 11.47   \\ 
 Hip 37217 & 07:38:41.0 & -21:13:28.5 & 0.29 & -0.27 & M3 & 11.7 & \cite{Koen10} & 10.60   \\
  Hip 37766 & 07:44:40.2 & +03:33:08.8 & 0.31 & 0.27 & M4.5 & 11.2 & \cite{Koen10} & 5.96  \\
    GJ 2066 & 08:16:08.0 & +01:18:09.3 & 0.46 & -0.10 & M2 & 10.1 & \cite{Koen10} & 9.12   \\ 
     Gl 317 & 08:40:59.2 & -23:27:23.3 & 0.43 & 0.20 & M3.5 &  12.0 & \cite{VanAltena95} &  15.31   \\ 
    HD 75732B & 08:52:40.8 & +28:18:59.0 & 0.27 & 0.15 & M4 & 13.1 & \cite{Gliese91} & 13.02  \\
  Hip 46655 & 09:30:44.6 & +00:19:21.6 & 0.29 & -0.17 & M3.5 & 11.7 & \cite{Koen10} & 9.67  \\
  Hip 46769 & 09:31:56.3 & +36.19:12.8 & 0.53 & -0.27 & M0 & 10.1 & \cite{Hog00} & 13.91   \\ 
     Gl 357 & 09:36:01.6 & -21:39:38.9 & 0.33 & -0.31 & M2.5 &  10.9 & \cite{Koen10} &  9.02  \\ 
  Hip 47513 & 09:41:10.4 & +13:12:34.4 &  0.48 & -0.12 & M1.5 & 10.4 & \cite{Koen10} & 11.26  \\ 
  Hip 47650 & 09:42:51.7 & +70:02:21.9 & 0.41 & 0.13 & M3 & 11.4 & \cite{Hog00} & 11.35    \\
  Hip 48714 & 09:56:08.7 & +62:47:18.5 & 0.64 & -0.03 & M0 &  9.00 & \cite{Gliese91} & 10.56   \\ 
     Gl 382 & 10:12:17.7 & -03:44:44.4 & 0.54 & 0.02 & M1.5 &  9.26 & \cite{Koen10} & 7.87  \\  
     Gl 388 & 10:19:36.3 & +19:52:10.1 & 0.41 & 0.10 & M3.5 & 9.46 & \cite{Hog00} & 4.69  \\
  Hip 51007 & 10:25:10.8 & -10:13:43.3 & 0.54 & -0.07 & M1 & 10.1 & \cite{Koen10} & 12.35 \\ 
     Gl 393 & 10:28:55.6 & +00:50:27.6 & 0.44 & -0.14 & M2 & 9.65 & \cite{Landolt09} &  7.07   \\
 Hip 53020 & 10:50:52.0 & +06:48:29.2 & 0.26 & 0.00 & M4 & 11.7 & \cite{Landolt92} & 6.76  \\
  Gl 406 & 10:56:28.9 & +07:00:52.8 & 0.10 & 0.22 & M5.5 & 13.5 & \cite{Landolt92} & 2.39   \\
     Gl 408 & 11:00:04.3 & +22:49:58.6 &  0.38 & -0.15 & M2.5 & 10.0 & \cite{Koen10} &  6.66 \\ 
    HD 95650 & 11:02:38.3 & +21:58:01.7 &  0.59 & -0.10 & M0 & 9.57 & \cite{Koen10} & 11.77   \\ 
    HD 95735 & 11:03:20.2 & +35.58:11.6 & 0.39 & -0.32 & M2 & 7.52 & \cite{Oja85} &  2.55  \\ 
  Hip 54532 & 11:09:31:3 & -24:35:55.1 & 0.46 & -0.08 & M2 & 10.4 & \cite{Koen10} & 10.75  \\ 
   HD 97101B & 11:11:01.9 & +30:26:44.4 & 0.58 & 0.52 & M1.5 & 10.7 & \cite{Hog00} & 11.87 \\
  Hip 55360 & 11:20:04.8 & +65:50:47.3 & 0.49 & -0.35 & M0 &  9.30 & \cite{Hog00} & 8.92  \\ 
     Gl 433 & 11:35:26.9 & -32:32:23.9 & 0.47 & -0.15 & M1.5 &  9.81 & \cite{Koen10} &  8.88    \\ 
  Hip 57050 & 11:41:44.6 & +42:45:07.1 & 0.35 & 0.08 & M4 & 11.9 & \cite{Kharchenko01} & 11.10  \\ 
  Gl 436 & 11:42:11.2 & +26:42:22.6 &  0.44 & -0.03 & M2.5 & 10.6 & \cite{Hog00} & 10.14   \\ 
     Gl 445 & 11:47:41.4 & +78:41:28.2 & 0.25 & -0.27 & M3.5 & 10.8 & \cite{Hog00} &  5.35  \\ 
 Hip 57548 & 11:47:44.4 & +00:48:16.4 & 0.17 & -0.23 & M4 & 11.1 & \cite{Landolt92} & 3.36 \\
     Gl 450 & 11:51:07.3 & +35:16:19.3 & 0.46 & -0.21 & M1 & 9.72 & \cite{Hog00} &  8.59  \\ 
  Hip 59406 & 12:11:11.8 & -19:57:38.1 & 0.35 & -0.13 & M3 & 11.7 & \cite{Koen10} & 12.59 \\ 
  Hip 59406b & 12:11:17.0 & -19:58:21.4 & 0.25 & -0.25 & M4 & 12.6 & \cite{Gliese91}  & 12.59  \\
  Hip 60559 & 12:24:52.5 & -18:14:32.2 & 0.26 & -0.56 & M4 & 11.3 & \cite{Koen10} & 8.85  \\
     Gl 486 & 12:47:56.6 & +09:45:05.0 & 0.32 & 0.01 & M3.5 &  11.4 & \cite{Koen10} &  8.37  \\ 
  Hip 63510 & 13:00:46.6 & +12:22:36.6 & 0.594 & 0.04 & M0.5 & 9.76 & \cite{Koen10} & 11.4  \\ 
     Gl 514 & 13:29:59.8 & +10:22:37.8 & 0.53 & -0.15 & M0.5 & 9.03 & \cite{Koen10} &  7.66  \\ 
   HD 119850 & 13:45:43.8 & +14:53:29.5 & 0.50 & -0.16 & M1.5 & 8.50 & \cite{VanBelle09} &  5.39 \\
  Hip 67164 & 13:45:50:7 & -17:58:05.6 & 0.31 & -0.06 & M3.5 & 11.9 & \cite{Koen10} & 10.24  \\ 
   HD 122303 & 14:01:03.2 & -02:39:17.5 & 0.52 & -0.16 & M1 & 9.71 & \cite{Koen10} & 10.03    \\ 
  Hip 70865 & 14:29:29.7 & +15:31:57.5 & 0.52 & 0.00 & M2 & 10.7 & \cite{Koen10} & 14.00 \\
  Hip 70975 & 14:31:01.2 & -12:17:45.9 & 0.32 & -0.05 & M3.5 & 11.9 & \cite{Koen10} & 10.82  \\ 
  Hip 71253 & 14:34:16.8 & -12:31:10.4 & 0.28 & 0.11 & M4 &  11.3 & \cite{Koen10} &  6.06 \\ 
  Hip 71898 & 14:42:21.6 & +66:03:20.9 & 0.361 & -0.35 & M3 & 10.8 & \cite{Hog00} &  9.87  \\ 
  Gl 569A & 14:54:29.2 & +16:06:03.8 & 0.48 & -0.03 & M2.5 & 10.2 & \cite{Koen10} & 9.65  \\
  Gl 581 & 15:19:27.5 & -07:43:19.4 & 0.30 & -0.18 & M3 & 10.6 & \cite{Hog00} & 6.21 \\
   HD 147379B & 16:16:45.3 & +67:15:22.5 & 0.47 & 0.09 & M3 & 10.7 & \cite{Gliese91} & 10.74 \\
     Gl 625 & 16:25:24.6 & +54:18:14.7 & 0.32 & -0.39 & M1.5 & 10.2 & \cite{Hog00} &  6.52  \\ 
  Gl 649 & 16:58:08.9 & +25:44:39.0 & 0.54 & -0.10 & M1 & 9.66 & \cite{Hog00} & 10.34  \\
  Hip 83762 & 17:07:07.5 & +21:33:14.5 & 0.38 & -0.10 & M3 & 11.7 & \cite{Koen10} & 13.4   \\
  Hip 84099 & 17:11:34.7 & +38:26:33.9 & 0.38 & -0.05 & M3.5 & 11.5 & \cite{Hog00} & 12.00   \\ 
  Hip 84790 & 17:19:52.7 & +41:42:49.7 & 0.37 & -0.21 & M2.5 & 11.4 & \cite{Gliese91} & 12.38  \\ 
     Gl 687 & 17:36:25.9 & +68:20:20.9 & 0.40 & -0.06 & M3 & 9.15 & \cite{Hog00} & 4.53  \\
     Gl 686 & 17:37:53.3 & +18:35:30.2 & 0.44 & -0.31 & M1 & 9.58 & \cite{Koen10} & 8.09   \\ 
     Gl 694 & 17:43:56.0 & +43:22:43.0 & 0.44 & -0.02 & M2.5 & 10.5 & \cite{Hog00} &  9.48   \\ 
     Gl 699 & 17:57:48.5 & +04:41:36.2 & 0.16 & -0.61 & M4 & 9.51 & \cite{Koen10} & 1.82   \\ 
   HD 165222 & 18:05:07.6 & -03:01:52.8 & 0.48 & -0.22 & M1 &  9.36 & \cite{Koen10} &  7.76 \\ 
 G205-028 & 18:31:58.4 & +40:41:10.4 & 0.31 & -0.14 & M3.5 & 12.0 & \cite{Gliese91} & 11.9 \\
    GJ 4063 & 18:34:36.6 & +40:07:26.4 & 0.19 & -0.61 & M3.5 & 11.8 & \cite{Hog00} &  7.25    \\ 
  Hip 91699 & 18:41:59.0 & +31:49:49.8 & 0.37 & -0.13 & M3 & 11.3 & \cite{Kharchenko01} & 11.45  \\
  Hip 92403 & 18:49:49.4 & -23:50:10.4 & 0.17 & -0.43 & M3.5 & 10.5 & \cite{Koen10} & 2.97  \\
    Gl 745A & 19:07:05.6 & +20:53:17.0 & 0.30 & -0.48 & M1.5 & 10.8 & \cite{Koen10} &  8.51  \\
    Gl 745B & 19:07:13.2 & +20:52:37.2 & 0.31 & -0.45 & M1.5 & 10.7 & \cite{Koen10} & 8.75  \\
 G207-019 & 19:08:30.0 & +32:16:52.0 & 0.34 & -0.10 & M3 & 11.8 & \cite{Kharchenko01} & 12.39   \\
   HD 180617 & 19:16:55.3 & +05:10:08.1 & 0.48 & 0.02 & M2.5 &  9.12 & \cite{Koen10} &  5.87  \\
     Gl 793 & 20:30:32.0 & +65:26:58.4 & 0.38 & -0.03 & M2.5 & 10.7 & \cite{Hog00} &  8.00  \\ 
     Gl 806 & 20:45:04.1 & +44:29.56.7 & 0.44 & -0.16 & M1.5 & 10.8 & \cite{Hog00} & 12.32    \\ 
 Hip 103039 & 20:52:33.0 & -16:58:29.0 & 0.23 & -0.10 & M4 & 11.4 & \cite{Koen10} & 5.71  \\
   HD 199305 & 20:53:19.8 & +62:09:15.8 &  0.58 & -0.02 & M0.5 &  8.60 & \cite{Hog00} &  7.05   \\ 
 Hip 104432 & 21:09:17.4 & -13:18:09.0 & 0.36 & -0.51 & M1 & 10.9 & \cite{Landolt09} & 12.17   \\ 
   HD 209290 & 22:02:10.3 & +01:24:00.8 & 0.60 & -0.10 & M0 & 9.15 & \cite{Koen10} & 10.24  \\ 
 Gl 849 & 22:09:40.3 & -04:38:26.6 & 0.49 & 0.22 & M3.5 & 10.4 & \cite{Koen10} &  9.10  \\
 Hip 109555 & 22:11:30.1 & +18:25:34.3 & 0.55 & 0.13 & M2 & 10.2 & \cite{Koen10} & 11.62   \\
     Gl 876 & 22:53:16.7 & -14:15:49.3 & 0.34 & 0.13 & M4 & 10.2 & \cite{Landolt09} &  4.69 \\
   HD 216899 & 22:56:34.8 & +16:33:12.4 & 0.58 & 0.03 & M1.5 & 8.64 & \cite{Koen10} & 6.84 \\
   HD 217987 & 23:05:52.0 & -35:51:11.0 & 0.47 & -0.33 & M0.5 & 7.34 & \cite{Gliese91} & 3.28  \\ 
 Hip 114411 & 23:10:15.7 & -25:55:52.7 & 0.46 & -0.13 & M2 & 11.3 & \cite{Koen10} & 16.08   \\ 
 Hip 115332 & 23:21:37.4 & +17:17:25.4 & 0.40 & 0.27 & M4 & 11.7 & \cite{Koen10} & 10.99  \\ 
 Hip 115562 & 23:24:30.5 & +57:51:15.5 & 0.59 & 0.08 &  M1 & 10.0 & \cite{Gliese91} & 12.96\\ 
 Gl 905 & 23:41:55.0 & +44:10:40.8 & 0.14 & 0.05 & M5 & 12.3 & \cite{Jenkins09} & 3.16 \\
     Gl 908 & 23:49:12.5 & +02:24:04.4 & 0.42 & -0.39 & M1 & 8.99 & \cite{Landolt09} &  5.98 
      \enddata
        \label{T1}
        \tablecomments{Metallicity uncertainties are taken to be 0.17 dex, while mass uncertainties are taken as $10\%$, following the method of \cite{Delfosse00}}
\end{deluxetable*}
\clearpage
r

\LongTables

\begin{deluxetable}{l|ccccccc}
        \tabletypesize{\scriptsize}
        \tablecolumns{8}
        \tablecaption{RV Observations}
        \tablehead{\colhead{Star} & \colhead{N$_\textrm{obs}$} & \colhead{Baseline (yr)} & \colhead{Med. $\sigma_{\gamma}$ (m s$^{-1}$)} & \colhead{Jitter (m s$^{-1}$)} & \colhead{RMS (m s$^{-1}$)} & \colhead{RV Planets} & \colhead{Binary Companion}}
        \startdata
Hip 428 &  41 &  12.2 &   1.6 &   4.2 &   4.8 & 0 & K6 \citep{Bidelman54}  \\ 
HD 225213  &  67 &   9.9 &   1.1 &   3.2 &   3.1 & 0 & -  \\ 
Hip 1734  &   8 &   8.1 &   2.6 &   4.7 &   7.6 & 0 & -  \\ 
Gl 26  &  40 &  11.6 &   2.8 &   2.9 &   7.7 & 0 & -  \\ 
Hip 3143  &   8 &   9.8 &   5.6 &   2.6 &  11.6 & 0 & -  \\ 
Gl 48  &  41 &  15.2 &   1.3 &   2.5 &   3.5 & 0 & -  \\ 
Gl 49  &  22 &  14.2 &   1.4 &   7.9 &   5.0 & 0 & -  \\ 
Hip 5643  &  15 &   7.1 &   3.4 &  13.2 &   7.8 & 0 & -  \\ 
Hip 8051  &  33 &  12.7 &   1.5 &   3.0 &   5.0 & 0 & -  \\ 
Gl 83.1  &  21 &   8.2 &   3.3 &  12.5 &  20.2 & 0 & -  \\ 
G244-047  &  10 &   7.5 &   2.8 &   3.6 &   4.0 & 0 & -  \\ 
Gl 87 &  62 &  13.0 &   1.3 &   2.5 &   7.4 & 0 & -  \\ 
Hip 11048  &  44 &  12.6 &   1.1 &   4.8 &   5.4 & 0 & -  \\ 
Gl 105B &  12 &   9.1 &   2.7 &   3.7 &  13.0 & 0 & K3 \citep{Gray06}  \\ 
Gl 109  &  32 &  13.1 &   1.4 &   2.8 &   4.4 & 0 & -  \\ 
Hip 21556  &  31 &  12.7 &   1.3 &   2.5 &   4.3 & 0 & -  \\ 
Gl 179 &  42 &  12.2 &   2.5 &   4.4 &  19.7 & 1 & -  \\ 
Hip 22762  &  39 &  12.6 &   1.6 &   2.7 &   4.6 & 0 & -  \\ 
Hip 23512  &  11 &   6.7 &   4.1 &   5.0 &   6.7 & 0 & -  \\ 
HD 33793  &  36 &  13.8 &   1.4 &   2.9 &   3.2 & 0 & -  \\ 
Hip 24284  &  30 &   9.1 &   1.4 &   2.3 &   5.4 & 0 & -  \\ 
HD 36395  &  33 &  15.8 &   1.7 &   5.7 &   7.8 & 0 & -  \\ 
G097-054  &  11 &   6.6 &   3.6 &   3.4 &   8.7 & 0 & -  \\ 
HD 233153 &  11 &   6.7 &   2.3 &   5.8 &   6.6 & 0 & K1 \citep{Montes01}  \\ 
Hip 26857  &  10 &   6.7 &   4.7 &   4.6 &  11.8 & 0 & -  \\ 
G192-13  &  16 &   7.8 &   4.3 &   4.1 &  11.4 & 0 & -  \\ 
Hip 29052  &  16 &   7.7 &   4.6 &   3.5 &  10.5 & 0 & -  \\ 
Gl 226  &  35 &  14.7 &   1.6 &   2.3 &  8.7 & 0 & -  \\ 
Gl 229B &  33 &  15.9 &   1.2 &   4.5 &   5.1 & 0 & T7 \citep{Faherty09}  \\ 
Gl 250B &  29 &   8.0 &   1.3 &   3.7 &   3.4 & 0 & K3 \citep{Gliese91}  \\ 
HD 265866  &  61 &  14.8 &   1.3 &   2.6 &   4.6 & 0 & -  \\ 
Gl 273  &  41 &  14.8 &   2.1 &   2.3 &  5.0 & 0 & -  \\ 
Hip 36338  &  10 &  10.7 &   2.9 &   2.3 &   5.8 & 0 & -  \\ 
Hip 36834  &  22 &   6.4 &   2.7 &   5.8 &  14.6 & 0 & -  \\ 
Hip 37217  &  11 &  11.8 &   3.4 &  25.7 &   5.3 & 0 & -  \\ 
Hip 37766  &  22 &  11.1 &   3.1 &  87.9 &  95.2 & 0 & -  \\ 
GJ 2066  &  37 &  14.8 &   1.5 &   2.5 &   5.3 & 0 & -  \\ 
Gl 317 &  45 &  12.1 &   2.2 &   4.5 &  56.9 & 1 & -  \\ 
HD 75732B &  21 &   9.1 &   5.2 &   4.9 &  17.1 & 0 & G8 \citep{Montes01}  \\ 
Hip 46655  &  11 &   6.0 &   3.9 &   2.9 &  18.6 & 0 & -  \\ 
Hip 46769  &  23 &   8.0 &   1.4 &   3.5 &   6.3 & 0 & -  \\ 
Gl 357  &  36 &  14.2 &   1.8 &   2.1 &   6.1 & 0 & -  \\ 
Hip 47513  &  29 &  12.1 &   1.4 &   3.8 &   6.1 & 0 & -  \\ 
Hip 47650  &  10 &   6.2 &   3.2 &  16.2 &  11.0 & 0 & -  \\ 
Hip 48714  &  16 &  11.2 &   1.4 &   6.3 &   9.6 & 0 & -  \\ 
Gl 382  &  29 &  12.9 &   1.5 &   5.3 &   6.4 & 0 & -  \\ 
Gl 388  &  39 &   5.7 &   1.8 &  24.0 &  17.9 & 0 & -  \\ 
Hip 51007  &  19 &  11.1 &   2.2 &   4.2 &   6.1 & 0 & -  \\ 
Gl 393  &  42 &  14.4 &   1.2 &   3.3 &   3.9 & 0 & -  \\ 
Hip 53020  &  12 &   6.3 &   3.4 &   6.5 &  13.0 & 0 & -  \\ 
Gl 406  &  21 &  13.0 &   6.8 & 20.1 &  15.0 & 0 & -  \\ 
Gl 408  &  39 &  14.8 &   1.4 &   3.1 &   4.2 & 0 & -  \\ 
HD 95650  &  30 &  11.1 &   1.8 &  10.8 &  14.8 & 0 & -  \\ 
HD 95735  & 211 &  15.2 &   1.0 &   2.7 &   3.9 & 0 & -  \\ 
Hip 54532  &  26 &  12.2 &   2.6 &   2.9 &  12.9 & 0 & -  \\ 
HD 97101B &  25 &  10.5 &   1.4 &   4.7 &   4.7 & 0 & K8 \citep{Gliese91}  \\ 
Hip 55360  &  30 &  11.9 &   2.4 &   2.2 &   8.2 & 0 & -  \\ 
Gl 433  &  27 &  13.1 &   2.4 &   2.4 &   6.8 & 0 & -  \\ 
Hip 57050 &  40 &  11.8 &   3.1 &   3.4 &  25.9 & 1 & -  \\ 
Gl 436 & 257 &  12.0 &   1.7 &   2.2 &  12.0 & 1 & -  \\ 
Gl 445  &  48 &  13.3 &   1.7 &   2.4 &   7.0 & 0 & -  \\ 
Hip 57548  &  17 &  12.8 &   2.8 &   9.2 &   5.9 & 0 & -  \\ 
Gl 450  &  31 &  14.1 &   2.0 &   4.7 &   7.0 & 0 & -  \\ 
Hip 59406 &  11 &   7.0 &   4.4 &   2.2 &  11.4 & 0 & M4 (Table \ref{T1}) \\ 
Hip 59406b &  12 &   6.2 &   6.1 &   3.2 &  13.2 & 0 & M3 (Table \ref{T1}) \\ 
Hip 60559  &  14 &   6.3 &   3.4 &   3.1 &   8.9 & 0 & -  \\ 
Gl 486  &  20 &   8.2 &   3.0 &   2.5 &  11.3 & 0 & -  \\ 
Hip 63510  &  41 &  11.3 &   3.4 &   6.0 & 1011.0 & 0 & M7 \citep{Beuzit04}  \\ 
Gl 514  &  50 &  13.9 &   1.4 &   3.5 &   6.0 & 0 & -  \\ 
HD 119850  &  42 &  13.9 &   1.3 &   2.2 &   3.2 & 0 & -  \\ 
Hip 67164  &  14 &   6.2 &   4.0 &   2.2 &   8.3 & 0 & -  \\ 
HD 122303  &  37 &  11.8 &   1.3 &   3.4 &   6.9 & 0 & -  \\ 
Hip 70865 &  21 &   8.5 &   1.8 &   2.7 &   7.5 & 0 & -  \\ 
Hip 70975  &  15 &  11.3 &   2.9 &   2.8 &   8.5 & 0 & -  \\ 
Hip 71253  &  21 &   7.9 &   2.7 &   4.2 &   8.1 & 0 & -  \\ 
Hip 71898  &  30 &  14.1 &   2.4 &   2.9 &  41.0 & 0 & L0 \citep{Faherty09}  \\ 
Gl 569A &  13 &   5.1 &   2.5 &  14.7 &   6.6 & 0 & M8.5+M9 \citep{Mason01}  \\ 
Gl 581 & 197 &  12.5 &   1.3 &   2.8 &   9.9 & 4 & -  \\ 
HD 147379B &  14 &   5.9 &   2.2 &   4.1 &   4.8 & 0 & M1 \citep{Herbig07}  \\ 
Gl 625  &  48 &  14.0 &   1.7 &   2.7 &   3.6 & 0 & -  \\ 
Gl 649 &  50 &  12.6 &   1.4 &   5.6 &   9.4 & 1 & -  \\ 
Hip 83762  &   8 &   2.9 &   1.3 &   2.8 &   7.1 & 0 & -  \\ 
Hip 84099  &  16 &   6.2 &   2.8 &   2.6 &   6.6 & 0 & -  \\ 
Hip 84790  &  17 &   4.9 &   3.0 &   2.2 &   5.6 & 0 & -  \\ 
Gl 687 & 100 &  13.8 &   1.2 &   2.3 &   5.9 & 0 & M3.5 \citep{Jenkins09}  \\ 
Gl 686  &  60 &  14.4 &   1.1 &   2.4 &   3.4 & 0 & -  \\ 
Gl 694  &  38 &  14.4 &   2.2 &   3.1 &   4.6 & 0 & -  \\ 
Gl 699  & 230 &  15.3 &   1.3 &   7.0 &   4.1 & 0 & -  \\ 
HD 165222  & 142 &  14.4 &   1.2 &   3.1 &   3.4 & 0 & -  \\ 
G205-028  &  12 &   6.2 &   3.8 &  27.6 &   8.1 & 0 & -  \\ 
GJ 4063  &  14 &   6.9 &   2.7 &   2.5 &   6.1 & 0 & -  \\ 
Hip 91699 &  17 &  12.0 &   2.9 &   3.4 &  11.6 & 0 & -  \\ 
Hip 92403  &  27 &   8.1 &   2.8 &   7.7 &  18.8 & 0 & -  \\ 
Gl 745A &  26 &  13.3 &   1.5 &   2.9 &   3.9 & 0 & M1.5 (Table \ref{T1})  \\ 
Gl 745B &  21 &  10.4 &   2.5 &   2.9 &   5.5 & 0 & M1.5 (Table \ref{T1})  \\ 
G207-019  &  12 &   6.2 &   3.3 &   9.7 &   7.9 & 0 & -  \\ 
HD 180617 & 143 &   9.8 &   1.3 &   3.3 &   4.7 & 0 & M8 \citep{Jenkins09}  \\ 
Gl 793  &  30 &  14.2 &   1.6 &   4.9 &  5.0 & 0 & -  \\ 
Gl 806  &  63 &  15.3 &   1.6 &   3.1 &   6.5 & 0 & -  \\ 
Hip 103039  &  19 &   8.2 &   3.4 &   5.5 &   6.7 & 0 & -  \\ 
HD 199305  &  45 &  15.3 &   1.1 &   4.5 &   3.3 & 0 & -  \\ 
Hip 104432  &  34 &  12.3 &   1.7 &   3.1 &   5.0 & 0 & -  \\ 
HD 209290  &  56 &  11.0 &   1.0 &   4.6 &   3.7 & 0 & -  \\ 
Gl 849 &  84 &  14.4 &   1.6 &   3.1 &  21.5 & 1 & -  \\ 
Hip 109555  &  16 &  11.1 &   2.5 &  12.5 &   8.4 & 0 & -  \\ 
Gl 876 & 207 &  14.4 &   2.1 &   4.0 & 150.4 & 4 & -  \\ 
HD 216899 &  50 &  15.1 &   1.1 &   4.2 &   4.6 & 0 & M2 \citep{Zakhozhaj02}  \\ 
HD 217987  &  69 &  14.3 &   1.2 &   3.3 &   4.9 & 0 & -  \\ 
Hip 114411  &  11 &   8.9 &   2.7 &   3.3 &   7.2 & 0 & -  \\ 
Hip 115332  &  14 &   6.7 &   3.4 &   3.2 &   9.2 & 0 & -  \\ 
Hip 115562  &  10 &   8.8 &   1.6 &   6.2 &   9.0 & 0 & -  \\ 
Gl 905  &  17 &   8.0 &   3.8 &  8.6 &   8.8 & 0 & -  \\ 
Gl 908  &  89 &  16.0 &   1.2 &   2.6 &   2.9 & 0 & -  \\ 
      \enddata
        \label{T1b}
\end{deluxetable}
\clearpage

\begin{deluxetable*}{l|cccc}[t!]
        \tabletypesize{\footnotesize}
        \tablecolumns{5}
        \tablecaption{Previously published RV planets}
        \tablehead{ \colhead{Star} & \colhead{Planet $m \sin i$ ($M_J$)} & \colhead{Period (days)} & \colhead{Discovery} & \colhead{Updated Parameters}}
        \startdata
     Gl\,179 & $0.82 \pm 0.07$ & $2288 \pm 59$ & \cite{Howard10} & \cite{Howard10} \\
     Gl\,317 & $1.80 \pm 0.05$ & $691.8 \pm 4.7$ & \cite{Johnson07} & \cite{Anglada-Escude12a}   \\
  Hip 57050 & $0.298 \pm 0.025$ & $41.397 \pm 0.016$ & \cite{Haghighipour10} & \cite{Haghighipour10} \\ 
  Gl 436 & $0.0737 \pm 0.0052$ & $2.643899 \pm 0.000001$ & \cite{Butler04} & \cite{Southworth10}  \\
 Gl 581 & $0.049 \pm 0.001$ & $5.369 \pm 0.002$ & \cite{Bonfils05} & \cite{TadeuDosSantos12} \\
        & $0.017 \pm 0.001$ & $12.931 \pm 0.002$ & \cite{Udry07} & \cite{TadeuDosSantos12} \\
        & $0.006 \pm 0.003$ & $1.0124 \pm 0.0001$ & \cite{Udry07} & \cite{TadeuDosSantos12} \\
        & $0.006 \pm 0.003$ & $2.149 \pm 0.002$ & \cite{Mayor09} & \cite{TadeuDosSantos12} \\
 Gl 649 & $0.328 \pm 0.032$ & $598.3 \pm 4.2$ & \cite{Johnson10b} & \cite{Johnson10b} \\
 Gl 849 & $0.82 \pm 0.07$ & $1890 \pm 130$ & \cite{Butler06} & \cite{Butler06}  \\ 
     Gl 876 & $1.9506 \pm 0.0039$ & $61.1166 \pm 0.0086$ & \cite{Marcy98} & \cite{Rivera10} \\ 
        & $0.612 \pm 0.003$ & $30.0881 \pm 0.0082$ & \cite{Marcy01} & \cite{Rivera10} \\
        & $0.018 \pm 0.001$  & $1.93778 \pm 0.00002$ & \cite{Rivera05}& \cite{Rivera10} \\
        & $0.039 \pm 0.005$  & $124.26 \pm 0.70$ & \cite{Rivera10}& \cite{Rivera10} 
        \enddata
        \label{knownpl}
\end{deluxetable*}

\begin{deluxetable*}{l|cccccc}[t!]
        \tabletypesize{\footnotesize}
        \tablecolumns{7}
        \tablecaption{Stars with measured RV accelerations and imaging nondetections}
        \tablehead{ \colhead{Star} & \colhead{RV Slope (m s$^{-1}$ yr$^{-1}$)} & \colhead{AO Obseration Date} & \colhead{Instrument} & \colhead{Filter} & \colhead{ADI} & \colhead{Cause of Acceleration}}
        \startdata
     Gl\,317 & $2.51 \pm 0.62$\tablenotemark{2} & 2010 October 13 & NIRC2  & $K^\prime$ & Yes & Presumed Companion  \\
  Gl\,179 & $-1.17 \pm 0.29$ &  2012 February 2 & NIRC2 &  $K^\prime$ & Yes & Presumed Companion \\ 
  Hip\,57050 & $1.39 \pm 0.39$ & 2012 December 27 & NIRC2 & $Ks$ &  No & Presumed Companion \\
 Gl\,849 & N/A\tablenotemark{1} & 2011 June 24 &  NIRC2 &  $L$ &  Yes & Identified Companion \\ 
  Hip\,63510 & N/A\tablenotemark{1} & N/A & N/A &  N/A & N/A & Brown Dwarf\tablenotemark{2}  \\ 
  Hip\,71898 & $8.6 \pm 0.4$ &  N/A & N/A &  N/A & N/A & Brown Dwarf\tablenotemark{3} \\

        \enddata
        \footnotetext[1]{Curvature in RV}\\        \footnotetext[2]{\cite{Beuzit04}}\\
        \footnotetext[3]{\cite{Golimowski04}}\\
        \label{T2}
\end{deluxetable*}

\clearpage

\appendix
\section{Notes on Individual Targets}
\label{Notes}

\subsection{Gl\,849}
The RV data for Gl\,849 exhibits a clear planetary signal from the known companion Gl\,849b. The residuals to the best-fitting orbit for this planet exhibit strong curvature, motivating our two-planet fit. Moreover, there is no correlation between this long period signal and stellar magnetic activity, suggesting the planet is not the result of an apparent velocity change during the star's magnetic cycle. To determine the orbital parameters of both planets, we utilize emcee, an affine invariant MCMC ensemble sampler \citep{Foreman-Mackey12}. For both planets, we fit five orbital parameters: the eccentricity $e$, argument of periapsis $\omega$, time at which a transit would occur $t_{\varpi=90}$, Doppler semiamplitude $K$ (or the product of the planet mass and the inclination $m \sin i$), and planet orbital period $P$. We also include the systemic radial velocity $\gamma$ as a free parameter, as well as a velocity offset between observations taken before August 18, 2004 and after that date, corresponding to an upgrade of the HIRES CCD detector \citep{Wright11}.

\begin{figure*}[htbp]
\centerline{\includegraphics[width=1.0\textwidth]{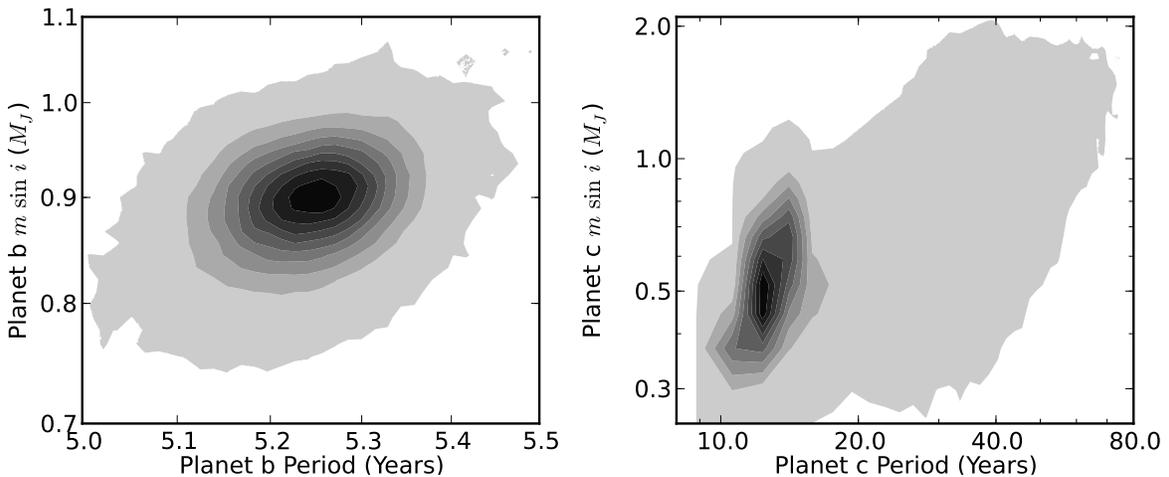}}
\caption{Position of (left) Gl\,849b and (right) Gl\,849c in the mass-period plane. The orbital parameters for the inner planet are much more tightly constrained than the outer planet. Depending on the exact shape of the planet distribution function, the inner planet may have more than a 50\% probability of being more massive than Jupiter when orientation uncertainties are taken into account.
  }
\label{849}
\end{figure*}

Due to the curvature in the outer planet's orbit, we are able to constrain the mass and period of both companions. As shown in Fig. \ref{849}, the orbit of the outer planet is only weakly constrained. Nevertheless, the data can rule out orbits with $m \sin i > 2.5 M_J$. Moreover, we refine the inner planet's parameters: we find the ``b'' component's best-fitting mass and period increase slightly, but the disributions for each are consistent with those found by \cite{Butler06}. Our parameters for each planet are included in Table \ref{T849}.

\begin{deluxetable*}{l|c|ccc}
        \tabletypesize{\footnotesize}
        \tablecolumns{2}
        \tablecaption{Orbital Parameters for Gl\,849}
        \tablehead{ Parameter & Mean & 50\% & 15.8\%\tablenotemark{2} & 84.2\%\tablenotemark{2}}
        \startdata
     \textit{Planet b} &   \\
  Orbital period \textit{P} (yr) & $5.241$ & 5.243 & -0.067 & +0.064  \\ 
  Planet mass\tablenotemark{1} \textit{$m \sin i$} ($M_J$) & .899 & 0.900 & -0.045 & +0.043 \\
  Time of potential transit \textit{t$_{\varpi=90}$} (JD-2440000) & 537.3 & 536.9 & -161.3 & +164.7 \\
  $e^{1/2} \cos{\omega}$ & -0.048 & -0.059 & -0.105 & +0.122 \\
  $e^{1/2} \sin{\omega}$ & 0.099 & 0.116 & -0.161 & +0.114 \\
     \textit{Planet c} & \\
  Orbital period \textit{P} (yr) & $24.04$ & 19.35 & -5.93 & +17.20  \\ 
  Planet mass\tablenotemark{1} \textit{$m \sin i$} ($M_J$) & 0.773 & 0.702 & -0.203 & +0.344 \\
  Time of potential transit \textit{t$_{\varpi=90}$} (JD-2440000) & 3586.3 & 5660.3 & -7356.0 & +2387.6\\
  $e^{1/2} \cos{\omega}$ & -0.311 & -0.346 & -0.185 & +0.260 \\
  $e^{1/2} \sin{\omega}$ & -0.348 & -0.361 & -0.234 & +0.253 \\
     \textit{System Parameters} & \\
  HIRES detector upgrade offset (m s$^{-1}$) & 17.07 & 17.18 & -5.25 & +5.01
        \enddata
        \footnotetext[1]{Assuming a stellar mass of 0.49 $M_{\odot}$}
        \footnotetext[2]{Values given relative to the 50\% data point.}
        \label{T849}
\end{deluxetable*}

\subsection{HIP\,109555}
When observing HIP\,109555 we detected a possible faint companion object located tens of arcseconds away. To prove this companion is not associated with the primary but is instead unrelated, we compare the proper motion of both objects by identifying them in the 2MASS catalog \citep{Skrutskie06} and the Palomar Observatory Sky Survey \citep{Abell59}. Comparing the POSS data collected 16 July 1950 to the 2MASS observation, we detect a proper motion for HIP\,109555 of 0.36 arcsec/yr, consistent with previously published results \citep{VanLeeuwen07}. The hypothetical companion motion, however, is only 5 milliarcseconds per year. Additionally, the companion is bluer in colors derived using the 2MASS J, H, and K filters than HIP\,109555. These are both consistent with the companion being a distant background object, and we neglect its presence in our analysis.

\subsection{HIP\,57050}
\label{HIP57050}
\begin{figure}[htbp]
\centerline{\includegraphics[width=0.5\textwidth]{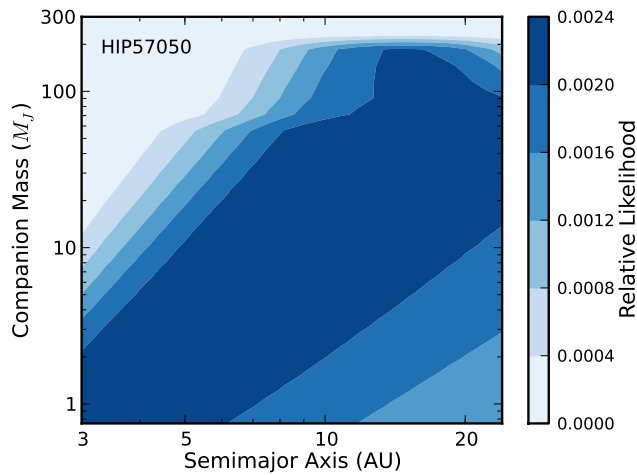}}
\caption{Probability contours displaying the location of a giant companion orbiting HIP\,57050, given that exactly one such planet exists, when the RV data is combined with adaptive optics imaging and 2MASS data. Because the AO imagery only extends to 11 AU, there is a small region of parameter space where a low-mass M-dwarf companion could reside. Additional AO observations with a wider field of view would be required to rule out this possibility. Lower-mass companions are allowed in shorter orbital periods due to possible curvature in the radial velocity data.
}
\label{Wright57050}
\end{figure}

We observed HIP\,57050 (=GJ1148) on December 27, 2012 using the $K_s$ filter on NIRC2. Our imaging is only complete at separations smaller than 1 arcsecond, corresponding to a projected separation of 11 AU. This does not enable us to rule out most stellar companions that could cause our observed RV trends, as shown in Fig. \ref{Wright57050}.
If the observed trend is caused by a stellar-mass companion, the companion is likely beyond 10 AU, which corresponds to a separation of 0.9 arcseconds. Thus any stellar companions at their maximum separation that could cause this trend would be expected to be found in a seeing-limited survey. We find no evidence for such a companion. While unlikely, additional AO observations with a wider field of view are required to fully eliminate the possibility that a low-mass star exists.

\subsection{HIP\,63510}
HIP\,63510B (Ross 458) is an M7 brown dwarf orbiting an M0.5 dwarf at approximately 3 AU \citep{Beuzit04}. Twelve years of RV observations suggest an orbit with a period of 13.9 years, an eccentricity of 0.32, and a minimum mass $m\sin i = 67.9 M_J$, suggesting a nearly edge-on orbit. We estimate a detection efficiency of 1.000 in an RV survey, which is not surprising considering the stellar RV semiamplitude is $K = 1.24$ km s$^{-1}$. This system contains a second companion which is separated from the host star by 1100 AU \citep{Goldman10, Scholz10}

\subsection{HIP\,71898}
\begin{figure}[htbp]
\centerline{\includegraphics[width=0.5\textwidth]{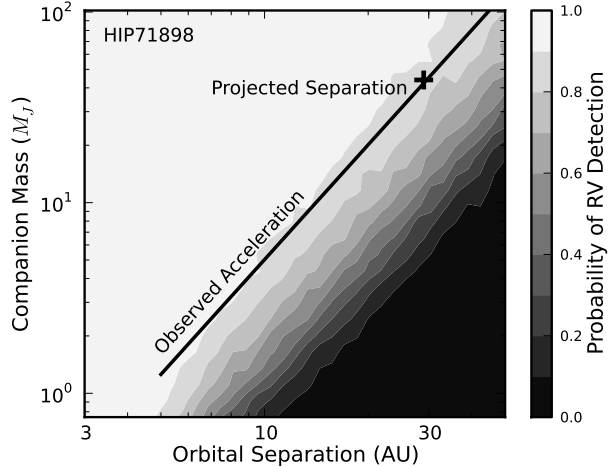}}
\caption{ Probability contours displaying the likelihood that a planet of a given mass and semimajor axis would be detected around HIP\,71898 in the CPS RV survey. The diagonal line represents companions that would produce an acceleration of $8.6 \pm 0.4$ m s$^{-1}$ yr$^{-1}$ in an edge on system when the companion was moving along the observer's line of sight. The $+$ marks the spot at which a $45 M_J$ companion at 30 AU would reside; this is the minimum mass and semimajor axis expected from this companion. 
  }
\label{71898}
\end{figure}

HIP\,71898B is an L0 dwarf in a wide orbit around an M3.5 dwarf. \cite{Golimowski04} report a projected separation of $30.01 \pm 3.78$ AU. This target has an RV baseline of 14 years, over which 30 observations were collected. From these observations we measure an acceleration of $8.6 \pm 0.4$ m s$^{-1}$ yr$^{-1}$. At 30 AU, this would suggest a minimum dynamical mass $m \sin i > 45 M_J$, consistent with an L0 dwarf. A detectability plot for companions to HIP\,71898 is shown in Fig. \ref{71898}. The observed acceleration lies near a contour representing a 0.9 probability of RV detection, so it is not surprising this companion was detected by CPS. 

\subsection{Gl569}
Gl\,569B is a brown dwarf binary, with an M8.5+M9 pair orbiting each other every $870 \pm 9$ days. The system has a combined mass of $0.140^{+0.009}_{-0.008} M_\odot$ \citep{Dupuy10} and is separated from the primary, an M3.5 dwarf, by a projected separation of 5 arcsec, or 47 AU \citep{Femenia11}. The maximum RV acceleration from such a companion is 3.7 m s$^{-1}$ yr$^{-1}$. For this star, we have a 5.1 year baseline and the median $\sigma$ is 15 m s$^{-1}$. By injecting simulated companions, we estimate an RV detection efficiency near zero for these companions. Thus it is not surprising that it is missed in our sample. 

\subsection{Gl\,229B}
Gl\,229B (HD\,42581) is a T7 dwarf at a projected separation of 44 AU \citep{Faherty09}. This companion has been directly imaged \citep{Nakajima95} but not detected as a strong acceleration through RV variations. As with Gl\,569, this object is beyond our range for efficient brown dwarf detection through RV observations. If we assume a mass of 40 $M_J$, we would expect a maximal RV acceleration of 1.1 m s$^{-1}$ yr$^{-1}$. Thus, again we should not be surprised it is not detected.

\section{A Brief Note on Radial Velocities and Magnetic Activity}
\begin{figure*}[htbp]
\centerline{\includegraphics[width=1.0\textwidth]{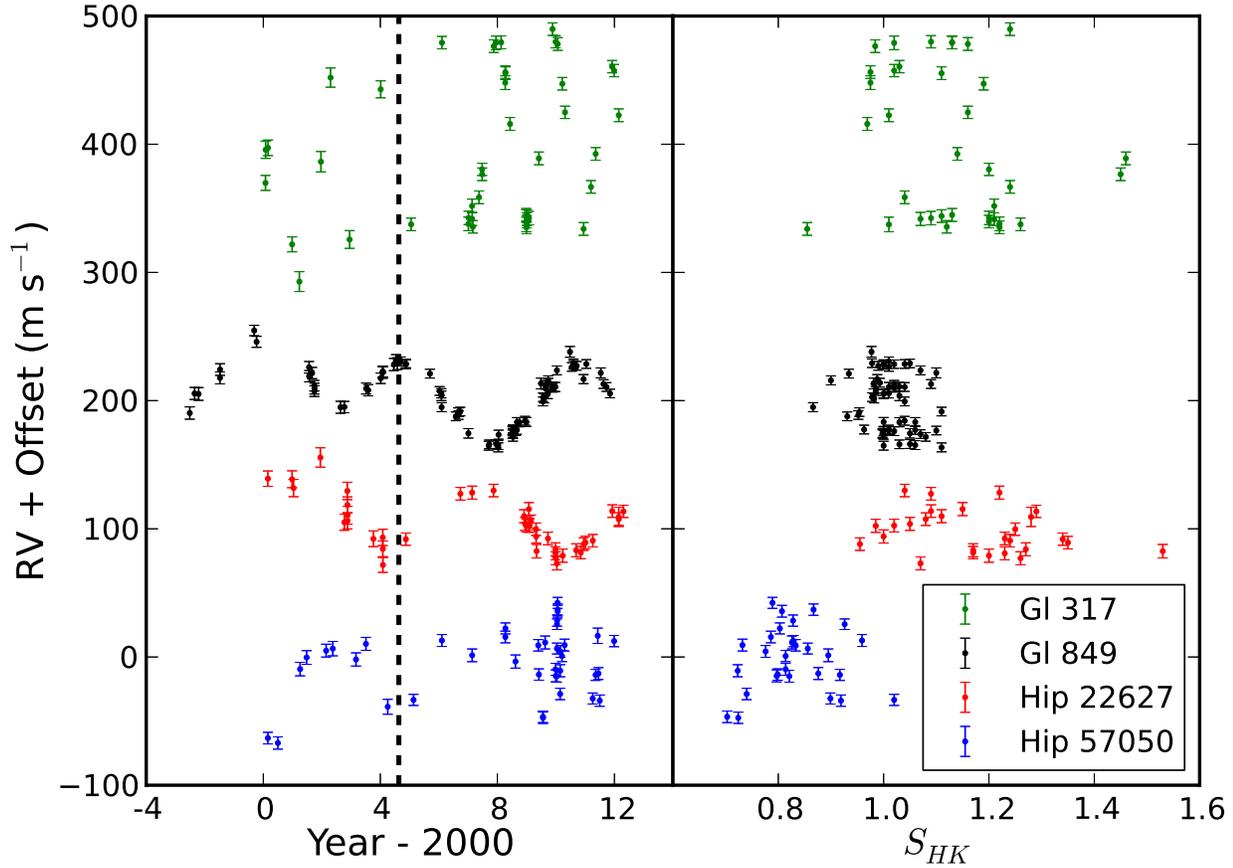}}
\caption{(left) RV time series for our four systems exhibiting long-term RV accelerations. The vertical line in 2004 represents the HIRES detector upgrade in August of that year. (right) RVs as a function of $S_\textrm{HK}$. All four RV accelerations are visible, but none of the RV data appear to correlate with $S_\textrm{HK}$, commonly used as a proxy for stellar chromospheric activity.
  }
\label{Trendplot}
\end{figure*}

We account for the possibility that any apparent RV accelerations may be induced by magnetic activity statistically, as described in \textsection\ref{FP}. Often, the $S_\textrm{HK}$ value, a measure of the ratio of flux in the Ca II line cores to flux in nearby continuum regions, is taken as a proxy for chromospheric activity \citep{Wilson68, Henry96}. While not a perfect measure, it is comforting to note that the observed radial velocities do not correlate with $S_\textrm{HK}$ in any of our stars with long-term RV accelerations. The RVs for our systems with detected accelerations as well as $S_\textrm{HK}$ for observations after the HIRES detector upgrade are included in Fig. \ref{Trendplot} and Table \ref{Txx}. \textbf{Note: We are open to this table being included as an online-only, machine readable table.}

\LongTables
\begin{deluxetable}{l|ccc||l|ccc}
        \tabletypesize{\scriptsize}
        \tablecolumns{8}
        \tablecaption{RVs and $S_{\textrm{HK}}$ values for systems with long-term RV accelerations.}
        \tablehead{\colhead{JD-244000} & \colhead{RV (m s$^{-1}$)} & \colhead{$\sigma_{\textrm{RV}}$ (m s$^{-1}$)} & \colhead{$S_{\textrm{HK}}$} & \colhead{JD-244000} & \colhead{RV (m s$^{-1}$)} & \colhead{$\sigma_{\textrm{RV}}$ (m s$^{-1}$} & \colhead{$S_{\textrm{HK}}$}}
        \startdata
Gl 317 & & & & & & & \\ 
11550.993 & 369.80 & 5.83 & N/A & 14544.905 & 456.22 & 4.92 & 0.97 \\  
11552.990 & 395.68 & 6.74 & N/A & 14545.894 & 455.37 & 4.96 & 1.11 \\  
11582.891 & 397.16 & 6.08 & N/A & 14603.777 & 415.82 & 5.00 & 0.97 \\  
11883.101 & 321.88 & 5.83 & N/A & 14806.029 & 344.83 & 5.02 & 1.13 \\  
11973.795 & 292.84 & 7.74 & N/A & 14807.069 & 337.39 & 5.70 & 1.01 \\  
12243.073 & 386.34 & 7.95 & N/A & 14808.138 & 343.93 & 4.94 & 1.11 \\  
12362.949 & 451.96 & 7.50 & N/A & 14809.059 & 335.15 & 4.95 & 1.22 \\  
12601.045 & 325.69 & 6.93 & N/A & 14810.161 & 339.65 & 4.87 & 1.20 \\  
12989.125 & 442.81 & 6.64 & N/A & 14811.128 & 341.51 & 4.93 & 1.21 \\  
13369.016 & 337.48 & 4.90 & 1.26 & 14839.107 & 342.36 & 5.26 & 1.09 \\  
13753.983 & 479.22 & 4.85 & 1.13 & 14963.795 & 388.95 & 4.98 & 1.46 \\  
14084.001 & 337.88 & 5.29 & 1.22 & 15134.090 & 489.75 & 4.90 & 1.24 \\  
14086.141 & 342.52 & 5.21 & 1.20 & 15173.079 & 479.92 & 4.79 & 1.09 \\  
14130.082 & 351.80 & 5.37 & 1.21 & 15199.017 & 478.18 & 4.98 & 1.16 \\  
14131.014 & 341.73 & 5.11 & 1.07 & 15255.869 & 447.15 & 4.89 & 1.19 \\  
14138.932 & 335.57 & 4.86 & 1.12 & 15289.857 & 424.86 & 4.82 & 1.16 \\  
14216.733 & 358.57 & 4.95 & 1.04 & 15522.057 & 333.93 & 4.97 & 0.85 \\  
14255.743 & 376.54 & 4.92 & 1.45 & 15613.960 & 366.70 & 4.92 & 1.24 \\  
14255.749 & 380.38 & 4.79 & 1.20 & 15672.848 & 392.45 & 4.92 & 1.14 \\  
14400.110 & 476.46 & 4.91 & 0.98 & 15878.127 & 460.49 & 4.81 & 1.03 \\  
14428.062 & 479.05 & 5.29 & 1.02 & 15903.017 & 457.41 & 4.79 & 1.02 \\  
14492.901 & 479.46 & 5.05 & 1.13 & 15960.986 & 422.57 & 4.98 & 1.01 \\  
14543.948 & 448.01 & 5.34 & 0.97 &  &  &  &  \\  
\hline
Gl 849 & & & & & & & \\ 
10606.068 & 190.31 & 4.78 & N/A & 14455.744 & 165.29 & 3.45 & 1.06 \\  
10666.001 & 205.60 & 4.69 & N/A & 14456.733 & 163.51 & 3.48 & 1.11 \\  
10715.957 & 205.19 & 4.99 & N/A & 14460.742 & 173.41 & 3.53 & 1.00 \\  
10983.038 & 217.69 & 4.67 & N/A & 14634.083 & 176.64 & 3.34 & 1.10 \\  
10984.084 & 224.23 & 4.55 & N/A & 14635.042 & 173.89 & 3.32 & 1.00 \\  
11410.021 & 254.67 & 4.08 & N/A & 14636.051 & 176.71 & 3.33 & 1.01 \\  
11439.865 & 245.85 & 4.30 & N/A & 14637.116 & 176.23 & 3.31 & 1.00 \\  
12095.081 & 225.97 & 4.52 & N/A & 14638.059 & 177.42 & 3.41 & 0.96 \\  
12096.046 & 219.06 & 4.38 & N/A & 14639.067 & 174.78 & 3.42 & 1.00 \\  
12133.013 & 221.49 & 4.39 & N/A & 14640.115 & 171.70 & 3.36 & 1.08 \\  
12160.909 & 211.60 & 4.10 & N/A & 14641.117 & 173.84 & 3.38 & 1.07 \\  
12161.846 & 207.39 & 4.19 & N/A & 14644.113 & 177.39 & 3.40 & 1.01 \\  
12162.887 & 209.34 & 4.22 & N/A & 14674.936 & 176.17 & 3.40 & 1.02 \\  
12486.968 & 194.80 & 4.66 & N/A & 14688.952 & 177.11 & 3.40 & 1.06 \\  
12535.852 & 194.96 & 4.43 & N/A & 14690.005 & 183.22 & 3.51 & 1.06 \\  
12807.011 & 209.44 & 4.30 & N/A & 14721.949 & 183.11 & 3.52 & 1.03 \\  
12834.013 & 208.07 & 4.39 & N/A & 14790.752 & 184.27 & 3.43 & 1.04 \\  
12989.720 & 217.41 & 4.08 & N/A & 14807.793 & 183.33 & 3.47 & 1.00 \\  
13014.710 & 222.75 & 4.27 & N/A & 14989.063 & 213.37 & 4.17 & 0.98 \\  
13015.711 & 221.97 & 4.60 & N/A & 15015.047 & 199.35 & 3.42 & 1.04 \\  
13016.706 & 222.33 & 4.07 & N/A & 15016.074 & 202.71 & 3.36 & 0.98 \\  
13154.080 & 228.16 & 4.76 & N/A & 15029.019 & 201.72 & 3.52 & 0.98 \\  
13180.108 & 231.43 & 4.45 & N/A & 15043.042 & 212.32 & 3.40 & 1.02 \\  
13196.931 & 228.82 & 4.63 & N/A & 15048.996 & 209.45 & 3.39 & 0.98 \\  
13238.929 & 230.55 & 3.44 & 1.01 & 15075.082 & 205.14 & 3.55 & 1.00 \\  
13301.838 & 228.44 & 3.39 & 1.00 & 15080.084 & 215.78 & 3.50 & 0.90 \\  
13302.742 & 228.98 & 3.32 & 1.05 & 15082.073 & 213.97 & 3.44 & 0.99 \\  
13303.798 & 228.40 & 3.27 & 1.02 & 15134.922 & 210.04 & 3.41 & 1.02 \\  
13603.939 & 221.04 & 3.43 & 0.93 & 15135.876 & 210.90 & 3.37 & 1.03 \\  
13724.712 & 207.52 & 3.39 & 0.98 & 15169.797 & 210.64 & 3.55 & 1.01 \\  
13746.715 & 205.70 & 3.60 & 1.01 & 15188.725 & 223.58 & 3.42 & 1.07 \\  
13746.721 & 203.74 & 3.72 & 1.03 & 15352.082 & 238.03 & 4.18 & 0.98 \\  
13749.698 & 194.88 & 3.51 & 0.87 & 15376.032 & 226.26 & 3.36 & 1.01 \\  
13927.015 & 187.71 & 3.42 & 0.93 & 15395.958 & 229.16 & 3.32 & 0.98 \\  
13959.087 & 191.03 & 3.34 & 1.90 & 15397.048 & 227.85 & 3.36 & 1.00 \\  
13960.955 & 188.72 & 3.31 & 0.95 & 15436.111 & 227.10 & 3.40 & 0.99 \\  
13960.962 & 191.05 & 3.32 & 0.95 & 15521.801 & 216.77 & 3.53 & 0.99 \\  
13983.000 & 191.46 & 3.36 & 1.11 & 15555.792 & 228.55 & 3.38 & 1.04 \\  
14083.750 & 174.45 & 3.67 & 1.05 & 15736.122 & 221.64 & 3.86 & 1.10 \\  
14337.074 & 164.82 & 3.45 & 1.00 & 15770.878 & 212.94 & 3.41 & 1.09 \\  
14343.872 & 165.90 & 3.35 & 1.03 & 15807.063 & 210.62 & 3.40 & 1.04 \\  
14429.742 & 166.12 & 3.44 & 1.05 & 15851.759 & 205.57 & 3.33 & 1.00 \\  
\hline
Hip 22627 & & & & & & & \\ 
11580.831 & 139.11 & 6.00 & N/A & 14838.995 & 115.36 & 5.10 & 1.15 \\  
11882.888 & 138.64 & 6.58 & N/A & 14846.957 & 102.80 & 5.28 & 2.81 \\  
11901.002 & 131.80 & 6.77 & N/A & 14864.957 & 105.69 & 5.05 & 1.97 \\  
12235.849 & 155.64 & 7.57 & N/A & 14928.732 & 99.68 & 4.86 & 1.25 \\  
12536.088 & 105.11 & 6.31 & N/A & 14929.726 & 94.02 & 5.14 & 1.00 \\  
12572.991 & 129.52 & 6.78 & N/A & 14934.731 & 82.53 & 5.28 & 1.53 \\  
12573.950 & 118.58 & 6.32 & N/A & 15077.110 & 92.45 & 4.91 & 1.23 \\  
12575.047 & 106.25 & 6.29 & N/A & 15170.784 & 80.88 & 5.08 & 1.23 \\  
12575.991 & 110.57 & 7.00 & N/A & 15170.791 & 84.01 & 5.07 & 1.27 \\  
12898.116 & 92.25 & 6.13 & N/A & 15174.093 & 77.01 & 5.12 & 1.26 \\  
13014.818 & 93.28 & 6.34 & N/A & 15187.837 & 72.99 & 5.07 & 1.07 \\  
13015.832 & 84.39 & 6.15 & N/A & 15261.771 & 79.05 & 5.15 & 1.20 \\  
13016.832 & 71.79 & 5.78 & N/A & 15429.120 & 83.25 & 5.07 & 1.17 \\  
13302.975 & 91.91 & 4.79 & 1.34 & 15487.096 & 81.47 & 4.81 & 1.17 \\  
13984.089 & 127.34 & 4.92 & 1.09 & 15522.938 & 88.05 & 4.89 & 0.95 \\  
14130.853 & 128.24 & 5.10 & 1.22 & 15545.819 & 89.25 & 4.89 & 1.35 \\  
14397.938 & 129.88 & 4.87 & 1.04 & 15636.775 & 90.74 & 4.87 & 1.24 \\  
14778.991 & 109.83 & 5.06 & 1.11 & 15879.984 & 113.81 & 4.90 & 1.09 \\  
14790.995 & 103.81 & 5.07 & 1.05 & 15960.761 & 109.29 & 7.51 & 1.28 \\  
14807.917 & 102.51 & 4.97 & 1.02 & 15960.765 & 107.53 & 4.93 & 1.08 \\  
14838.988 & 102.33 & 5.11 & 0.98 & 16019.733 & 113.58 & 4.78 & 1.29 \\  
\hline
Hip 57050 & & & & & & & \\ 
11581.046 & -63.25 & 4.53 & N/A & 15172.138 & -9.58 & 4.64 & 0.81 \\  
11705.827 & -67.09 & 4.79 & N/A & 15174.138 & -14.72 & 4.67 & 0.80 \\  
11983.009 & -9.42 & 5.27 & N/A & 15188.151 & -14.99 & 4.64 & 0.82 \\  
12064.864 & -0.39 & 5.34 & N/A & 15189.155 & 6.60 & 4.25 & 0.86 \\  
12308.077 & 4.98 & 5.01 & N/A & 15190.153 & 25.56 & 4.11 & 0.93 \\  
12391.034 & 6.53 & 5.63 & N/A & 15191.133 & 28.40 & 4.36 & 0.83 \\  
12681.050 & -1.92 & 5.15 & N/A & 15197.136 & 42.23 & 4.28 & 0.79 \\  
12804.885 & 10.26 & 5.05 & N/A & 15198.054 & 35.70 & 4.59 & 0.81 \\  
13077.104 & -38.83 & 5.83 & N/A & 15199.170 & 36.95 & 4.42 & 0.87 \\  
13398.975 & -33.44 & 4.33 & 1.02 & 15229.114 & -28.84 & 4.45 & 0.74 \\  
13753.068 & 12.88 & 4.64 & 0.96 & 15229.958 & -10.72 & 4.71 & 0.72 \\  
14131.092 & 1.32 & 4.96 & 0.90 & 15232.054 & 4.35 & 4.63 & 0.78 \\  
14545.002 & 15.61 & 4.55 & 0.79 & 15251.997 & 0.76 & 4.38 & 0.81 \\  
14546.007 & 22.29 & 4.29 & 0.80 & 15284.858 & 9.29 & 4.64 & 0.73 \\  
14671.811 & -3.54 & 5.13 & 5.32 & 15636.023 & -32.36 & 4.31 & 0.90 \\  
14955.894 & 9.12 & 4.47 & 0.83 & 15671.915 & -13.93 & 4.34 & 0.92 \\  
14963.930 & -13.66 & 4.43 & 0.80 & 15698.820 & 16.56 & 5.97 & 7.95 \\  
15014.782 & -46.65 & 4.50 & 0.70 & 15707.812 & -12.89 & 4.66 & 0.88 \\  
15015.804 & -47.40 & 4.41 & 0.72 & 15723.769 & -34.04 & 4.41 & 0.92 \\  
15041.758 & 11.31 & 5.04 & 0.83 & 15903.064 & 12.42 & 4.40 & 0.83  
        \enddata
        \label{Txx}
\end{deluxetable}

\end{document}